\documentclass[aps,prb,twocolumn,amsfonts,superscriptaddress]{revtex4-1}

\usepackage[latin1]{inputenc}
\usepackage{amsmath,amssymb}
\usepackage{graphicx,color}
\usepackage{braket}
\usepackage{subfigure}
\usepackage{bm}

\usepackage{titlesec}

\usepackage[breaklinks]{hyperref}
\usepackage[dvipsnames]{xcolor}
\hypersetup{colorlinks,citecolor=blue,filecolor=blue,linkcolor=blue,urlcolor=blue}

\usepackage{natbib}
\setcitestyle{square,numbers}

\newcommand{\ie}[0]{\textit{i.e. }}

\newcommand{\mc}[1]{\mathcal{#1}}
\newcommand{\mb}[1]{\mathbb{#1}}
\newcommand{\trm}[1]{\textrm{#1}}

\newcommand{\tbf}[1]{\textbf{#1}}
\renewcommand{\vec}[1]{\bm{#1}}

\allowdisplaybreaks[0]

\begin{document}

\title{Quadrupole spin polarization as  signature of  second-order  topological superconductors}

\author{Kirill Plekhanov}
\affiliation{Department of Physics, University of Basel, Klingelbergstrasse 82, CH-4056 Basel, Switzerland}

\author{Niclas M\"{u}ller}
\affiliation{Institut f\"{u}r Theorie der Statistischen Physik, RWTH Aachen, 52056 Aachen, Germany and JARA - Fundamentals of Future Information Technology}

\author{Yanick Volpez}
\affiliation{Department of Physics, University of Basel, Klingelbergstrasse 82, CH-4056 Basel, Switzerland}

\author{Dante M. Kennes}
\affiliation{Institut f\"{u}r Theorie der Statistischen Physik, RWTH Aachen, 52056 Aachen, Germany and JARA - Fundamentals of Future Information Technology}
\affiliation{Max Planck Institute for the Structure and Dynamics of Matter and Center for Free-Electron Laser Science, 22761 Hamburg, Germany}

\author{Herbert Schoeller}
\affiliation{Institut f\"{u}r Theorie der Statistischen Physik, RWTH Aachen, 52056 Aachen, Germany and JARA - Fundamentals of Future Information Technology}

\author{Daniel Loss}
\affiliation{Department of Physics, University of Basel, Klingelbergstrasse 82, CH-4056 Basel, Switzerland}

\author{Jelena Klinovaja}
\affiliation{Department of Physics, University of Basel, Klingelbergstrasse 82, CH-4056 Basel, Switzerland}

\date{\today}

\begin{abstract}
  We study theoretically second-order topological superconductors
  characterized by the presence of pairs of zero-energy Majorana
  corner states.  We uncover a quadrupole spin polarization at the
  system edges that provides a striking signature to identify
  topological phases, thereby complementing standard approaches based
  on zero-bias conductance peaks due to Majorana corner states.  We
  consider two different classes of second-order topological
  superconductors with broken time-reversal symmetry and show that
  both classes are characterized by a quadrupolar structure of the
  spin polarization that disappears as the system passes through the
  topological phase transition. This feature can be accessed
  experimentally using spin-polarized scanning tunneling
  microscopes. We study different models hosting second-order
  topological phases, both analytically and numerically, and using
  Keldysh techniques we provide numerical simulations of the
  spin-polarized currents probed by scanning tips.
\end{abstract}

\maketitle
\renewcommand{\thesection}{\Roman{section}}
\titleformat{\section}[runin]{\normalfont\itshape}{\thesection.}{3pt}{}[.]

{\it Introduction.} Over the last two decades, topological insulators
(TIs) and superconductors (TSCs) have become a subject of wide
interest in condensed matter physics~\cite{TI_KaneMele2005_1,
  TI_KaneMele2005_2, TI_BernevigHughesZhang2006, TIRev_HasanKane2010,
  TIRev_QiZhang2011, TIRev_SatoAndo2017, TIRev_WangZhang2017,
  TIRev_Wen2017}. One of the main attractions of such systems is the
existence of topologically protected gapless $(d-1)$-dimensional modes
which emerge at the boundary of a topologically nontrivial
$d$-dimensional bulk -- a phenomenon known as bulk-boundary
correspondence. The topological nature of the boundary modes makes
them insensitive to external perturbations and disorder, which is of
great importance in the context of quantum
computing~\cite{QComp_FreedmanKitaevLarsenWang2003, QComp_Kitaev2003,
  QComp_SternLindner2013}. Recently, the concept of bulk-boundary
correspondence has been generalized to a new class of systems, called
higher-order topological insulators and
superconductors~\cite{HOTI_BenalcazarBernevigHughes2017,
  HOTI_BenalcazarBernevigHughes2017_2, HOTI_SongFangFang2017,
  HOTI_LangbehnPengTrifunovicEtAl2017}. In contrast to conventional
topological systems, the $(d-1)$-dimensional boundary of an $n$-th
order TI/TSC is gapped. Instead, it exhibits protected gapless modes
on $(d-n)$-dimensional boundaries. The corresponding gapless modes are
called corner states in the case $n=d\geq2$.

Pioneering theoretical works on higher-order TIs have been followed by
fast progress from the experimental
side~\cite{PetersonBenalcazarHughesBahl2018, ImhofEtAl2018,
  MittalEtAl2019, ChenEtAl2019, HassanKunstEtAl2019,
  SerraGarciaEtAl2018, XueYangEtAl2019, SchindlerEtAl2018_1,
  SchindlerEtAl2018_2, WeinerNiLiAluKhanikaev2019,
  ZhangXieEtAl2019}. Particular attention has been dedicated to
theoretical investigations of TSCs that host Majorana corner states
(MCSs) -- Majorana bound states (MBSs) located at the corners of the
system~\cite{ShiozakiSato_2014, GeierEtAl_2018, Khalaf_2018,
  Trifunovic_2020, Zhu_2018, VopezEtAl_2019, LaubscherEtAl_2019,
  AhnYang_2020, WangLinHughes_2018, LiuHeNori_2018,
  FrancaEfremovFulga_2019, Yan_2019, ZhangColeWeDasSarma_2019,
  Zhu_2019, LaubscherEtAl_2020, WuLiuThomaleLiu_2019, WuEtAl_2019,
  ZhangCalzonaTrauzettel_2020, kramers_YanSongWang_2018,
  kramers_WangLiuLuZhang_2018, kramers_HsuEtAl_2018,
  kramers_ChenLiuXuLiu_2019,
  kramers_HsuColeZhangSau_2019}. Nevertheless, the experimental
realization of such systems remains challenging. Usual protocols to
detect MBSs in (higher-order) TSCs are based on a direct state
tomography or detection of specific features, such as a zero-bias peak
in the differential conductance. However, such probes do not provide a
clear way to distinguish between MBSs and other types of bound states
of topologically trivial nature~\cite{ZitkoLimLopezAquado2015,
  PradaEtAl2020, AlspaughSheehyGoerbigSimon2020,
  ZhangTrauzettel_2020,abs,abs1,abs2,abs3,abs4,abs5,abs6,abs7,abs8,abs9,sch},
which hinders unambiguous identification of the topological phases and
calls for additional experimental signatures.

In this work, we propose a solution to the problem described above,
based on an alternative probe of second-order TSCs (SOTSCs) with
broken time-reversal symmetry, which can be implemented with the help
of scanning tunneling microscopes (STMs)~\cite{SzumniakEtAl2017,
  SerinaLossKlinovaja2018, ThakurathiEtAl2020, MullerEtAl2020,
  JozwiakEtAl2016, JeonEtAl2017, JackEtAl2019, KimEtAl2019,
  RubyAl2019}.  We consider a subclass of SOTSCs represented by two
simple models: the first one supports two MCSs at a single pair of two
opposite corners and the second one supports four MCSs--one at every corner
of the setup. The topological phase transition in such systems is
accompanied by a drastic change in the spin polarization structure at
the system edges, which we denote as a quadrupolar structure of
the spin polarization. The analytical arguments justifying the
emergence of such edge features are confirmed numerically. We
then provide results of numerical simulations of the current flowing
between a local spin-polarized probe (an STM tip) and the sample,
making use of Keldysh techniques. We expect that our approach can also
be applied to probe other higher-order topological phases with
broken time-reversal symmetry. Importantly, the proposed signatures
are stable against weak disorder and do not rely on any other symmetry
than the particle-hole symmetry of the superconductor.

\begin{figure}[t]
  \centering
  \includegraphics[width=.99\columnwidth]{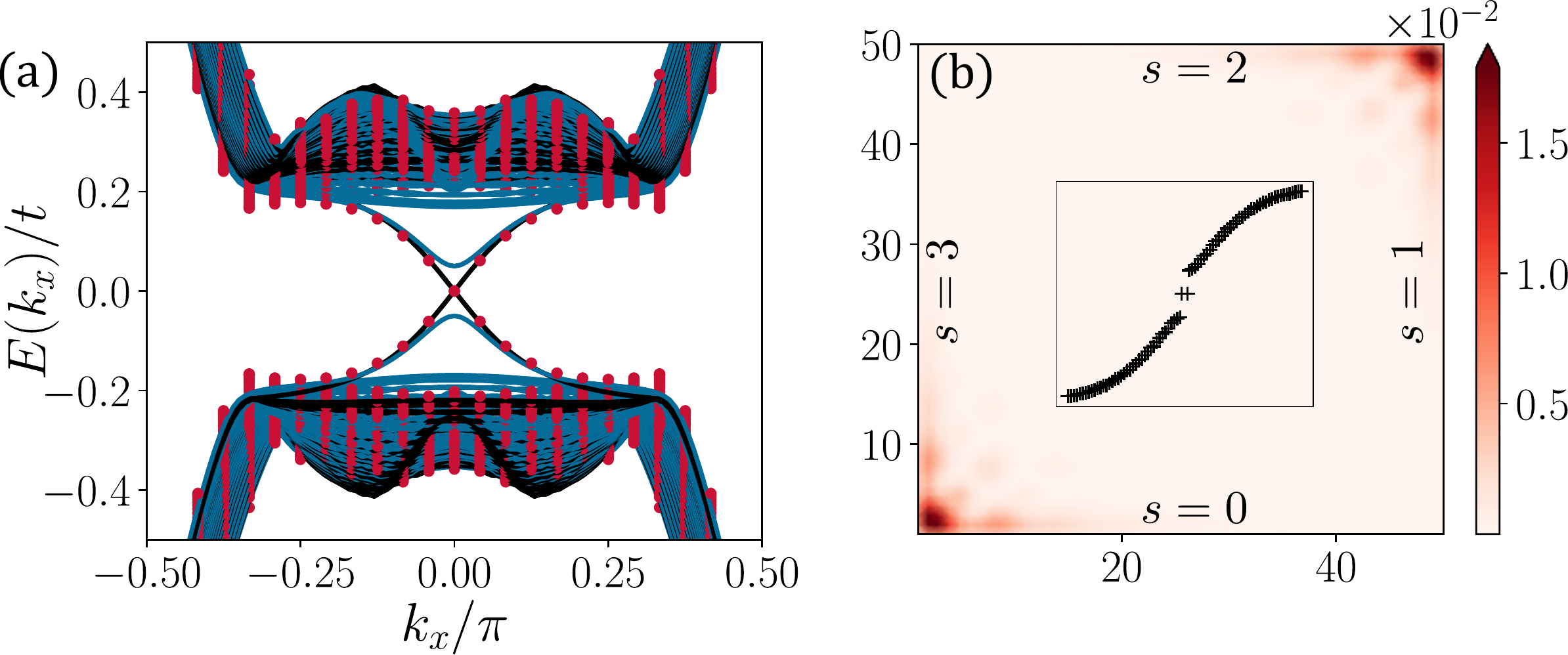}
  \caption{(a) Spectrum of $\mc{H}$ calculated in a geometry with OBC
    along the $y$ axis for $\Delta_{\trm{z}} = 0$ (black lines),
    $\Delta_{\trm{z}} = 0.05 t$ and $\theta_{\trm{z}} = 0$ (blue
    lines), as well as $\Delta_{\trm{z}} = 0.05 t$ and
    $\theta_{\trm{z}} = \pi / 2$ (red dots). We see that only the $x$
    component of the Zeeman field gaps out the helical edge modes,
    which are present in the system when $\Delta_{\trm{z}} = 0$. (b)
    Probability density of the MCSs calculated in a geometry with OBCs
    along both $x$ and $y$ axis for $\Delta_{\trm{z}} = 0.05t$ and
    $\theta_{\trm{z}} = \pi / 4$. The inset shows the low-energy
    spectrum. All the simulations are performed in the topological
    regime of the phase diagram with
    $\Gamma = 2 \Delta_{\trm{sc}} = \alpha = 0.5 t$, $\mu_0 = 0$, and
    $a = 1$.}
  \label{fig:soti_inversion}
\end{figure}

{\it SOTSCs with two corner states.}  To begin with, we consider a
SOTSC which supports a single pair of corner
states~\cite{ShiozakiSato_2014, GeierEtAl_2018, Khalaf_2018,
  Trifunovic_2020, Zhu_2018, VopezEtAl_2019, LaubscherEtAl_2019,
  AhnYang_2020}. The starting point of our consideration is a simple
2D model for a helical TSC~\cite{VopezEtAl_2019}, described by the
following Hamiltonian in momentum representation:
\begin{align}
  \label{eq:h0_inversion}
  & \mc{H}_0({\vec{k}}) = 2t
    \left[ 2 - \cos(k_x a) - \cos(k_y a) \right] \eta_z - \mu_0 \eta_z +
    \Gamma \eta_z \tau_x
    \notag \\
  & +   \alpha \left[ \sin(k_y a) \sigma_x - \sin(k_x a) \eta_z \sigma_y \right] \tau_z +
    \Delta_{\trm{sc}} \eta_y \sigma_y \tau_z .
\end{align}
The Pauli matrices $\eta_j$ act on the particle-hole space, $\sigma_j$
-- on the spin space, and $\tau_j$ -- on a generic local degree of
freedom (e.g. an electron orbital). We work in the Nambu basis
$(\psi_{\uparrow 1}, \psi_{\downarrow 1}, \psi^{\dag}_{\uparrow 1},
\psi^{\dag}_{\downarrow 1}, \psi_{\uparrow \bar{1}}, \psi_{\downarrow
  \bar{1}}, \psi^{\dag}_{\uparrow \bar{1}}, \psi^{\dag}_{\downarrow
  \bar{1}})$, where $\psi^{\dag}_{\sigma \tau}$ creates an electron of
species $\tau$ and spin $\sigma$. The parameters $\Delta_{\trm{sc}}$
and $\mu_0$ describe the $s$-wave superconducting pairing amplitude
and the chemical potential, respectively, while  $t$, $\alpha $ as well as
$\Gamma$ depend on the microscopic details of the system and $a$ is the lattice constant (see
Supplemental Material (SM)~\cite{SM} for more details).  The proposed model is characterized by a topological phase
transition at $\Gamma = \Delta_{\trm{sc}}$. The region
$\Gamma < \Delta_{\trm{sc}}$ is trivial, while
$\Gamma > \Delta_{\trm{sc}}$ corresponds to a helical TSC, supporting
a pair of gapless edge modes. The existence of these edge modes is
protected by the time-reversal symmetry $T = i \sigma_y K$, obeying
$T \mc{H}_0(\vec{k}) T^{-1} = \mc{H}_0(-\vec{k})$, with $K$ being the
complex conjugation operator. The topological phase diagram can be
checked numerically in a geometry with open boundary condition (OBC)
along one fixed direction [see Fig.~\ref{fig:soti_inversion}(a)]. The
resulting spectrum is independent of the particular choice of the OBC
direction as a result of the in-plane rotational symmetry present in
the system.

The helical edge modes can be gapped out by applying an external Zeeman field,
which breaks time-reversal symmetry. The corresponding contribution to the Hamiltonian can be expressed as
\begin{align}
  \label{eq:hz_inversion}
  \mc{H}_{\trm{z}}
  = \Delta_{\trm{z}} \left[
  \cos(\theta_{\trm{z}})\eta_z \sigma_x + \sin(\theta_{\trm{z}})\sigma_y
  \right] ,
\end{align}
where $\Delta_\trm{z}$ ($\theta_{\trm{z}}$) defines the strength
(in-plane orientation) of the Zeeman field. The total Hamiltonian then
becomes $\mc{H}(\vec{k}) = \mc{H}_0({\vec{k}}) +
\mc{H}_{\trm{z}}$. One can easily show both numerically and
analytically that edge modes are gapped out only by the Zeeman field
component that is parallel to the edge, see
Fig.~\ref{fig:soti_inversion}(a). As a result, the rotational symmetry
is broken down to the inversion symmetry $I = \eta_z \tau_x$, which
satisfies $I \mc{H}(\vec{k}) I^{-1} = \mc{H}(-\vec{k})$. When the
Zeeman field is smaller than the bulk gap of the TSC,
$\Delta_{\trm{z}} \ll \left| \Gamma - \Delta_{\trm{sc}} \right|$, the
boundary physics of the system away from the corners is described by a
$2 \times 2$ Jackiw-Rebbi Hamiltonian~\cite{Jackiw1Rebbi976,
  JackiwSchrieffer1981}
\begin{align}
  \label{eq:heff_inversion}
  \mc{H}_{\trm{eff}, s}(k_{s})
  = v_{\trm{F}} k_{s} \rho_z - m_{s} \rho_y ,
\end{align}
where $\rho_j$ act on the space of helical states
$\Ket{\Psi^{s}_{0,\pm}}$ belonging to the edge
$s \in \lbrace 0, 1, 2, 3 \rbrace$ [see
Fig.~\ref{fig:soti_inversion}(b)], $k_{s}$ denotes the momentum
parallel to the edge, $v_{\trm{F}}$ the Fermi velocity. The strength
of the mass term is given by
$m_s = \Delta_{\trm{z}} \cos(\theta_{\trm{z}} - \theta_s)$ with
$\theta_s = s \pi / 2$ (see SM~\cite{SM}). We denote by
$\Ket{\Psi^{s}_{0,+}}$ ($\Ket{\Psi^{s}_{0,-}}$) the states which move
clockwise (anticlockwise). Most importantly, as a result of the
presence of the inversion symmetry, opposite edges are necessarily
described by opposite signs of the mass term $m_{s}$ (\ie $m_0 = -m_2$
and $m_1 = - m_3$). Consequently, in a finite-size geometry where
$m_s$ is finite on every edge, there exist two corners connecting two
edges with opposite signs of $m_{s}$. Such gap inversion corners host
zero-energy states identified with MCSs of a SOTSC. Numerical evidence
of the existence of MCSs is shown in
Fig.~\ref{fig:soti_inversion}(b). We note that such a SOTSC phase
remains stable against arbitrary types of disorder as long as
additional perturbations do not close the surface
gap~\cite{GeierEtAl_2018, Khalaf_2018, Trifunovic_2020}.

\begin{figure}[t]
  \centering
  \includegraphics[width=.99\columnwidth]{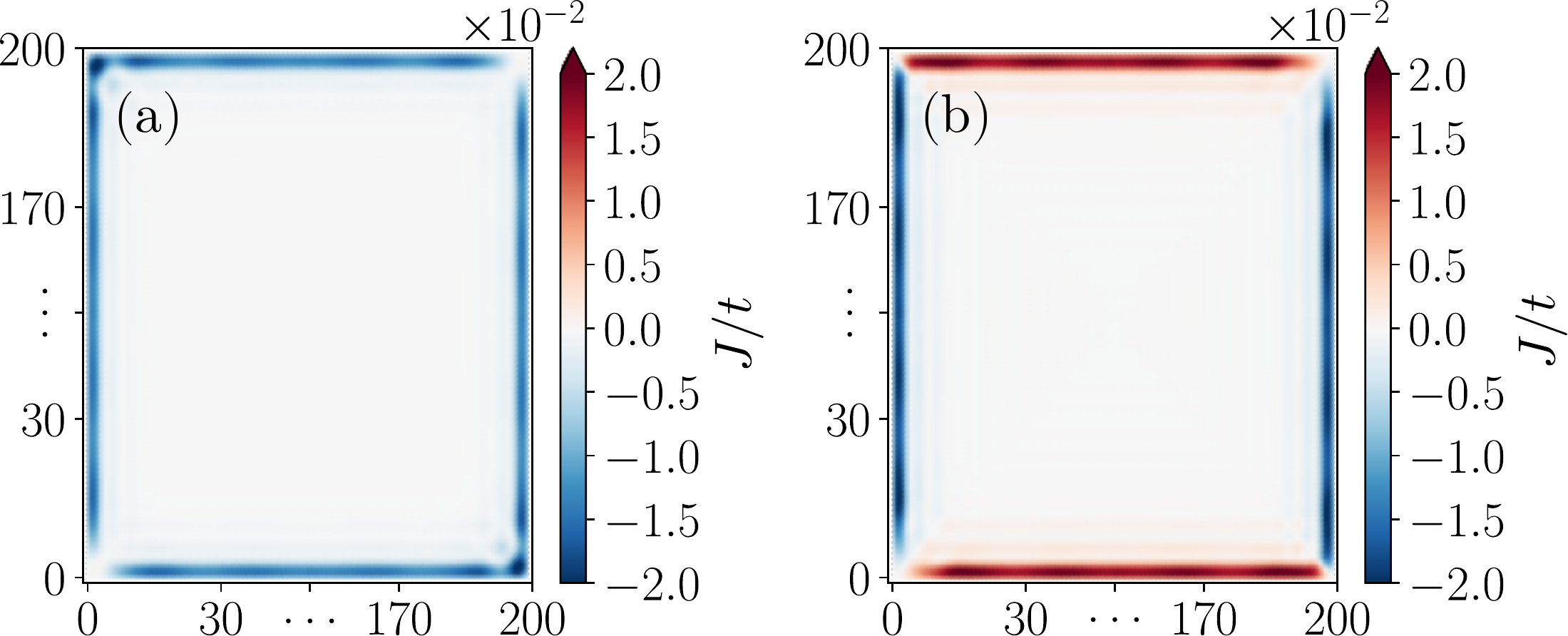}
  \caption{Numerical calculations of the current $J$ through the
    spin-polarized STM tip for a system described by $\mc{H}$ given
    below Eq.~(\ref{eq:hz_inversion}). The polarization of the STM is
    chosen to be in-plane (a) parallel to the Zeeman field and (b)
    perpendicular to the Zeeman field. The out-of-plane component is
    trivially zero. We see that while the parallel component of the
    polarization is roughly constant and negative along the entire
    boundary, the perpendicular component has a quadrupolar structure,
    \ie changes sign on every edge. The parameters of the simulation
    are $\Gamma = 1.75 \Delta_{\trm{sc}} = 0.44 t$,
    $\Delta_{\trm{z}} = 0.05 t$, $\kappa = 2.5 \cdot 10^{-4} t$, and
    the contribution to the current is summed over $V$ from $0$ to $\Delta_{\trm{z}}$.}
  \label{fig:soti_inversion_stm}
\end{figure}

{\it Quadrupolar structure of $\Braket{S_{\perp}}$.}
At low energies, the physics of the setup is dictated by the Zeeman
term. In particular, the eigenstates of $\mc{H}_{\trm{eff},s}$ at
$k_s = 0$ become spin-polarized and acquire the form
\begin{align}
  \Ket{\Psi^{s}_{\pm}} = \left( \Ket{\Psi^{s}_{0,+}} \mp
i \  \trm{sgn}(m_s) \Ket{\Psi^{s}_{0,-}} \right) / \sqrt{2}.
\end{align}
The states $\Ket{\Psi^{s}_{\pm}}$ correspond to the eigenvalues
$\pm |m_s|$ and have the spin polarization
$\Braket{\Psi^{s}_{\pm} | S_{\parallel} | \Psi^{s}_{\pm}} = \pm |m_s|
/ \Delta_{\trm{z}}$, where $S_{\parallel}$ is the spin operator along
the direction of the Zeeman field.  Similarly, one can calculate the
expectation values of the in-plane polarization perpendicular to the
applied Zeeman field, associated with the operator $S_{\perp}$. In our
case, we use the rotational symmetry of $\mc{H}_0$ and notice that the
gapless states at the edge $s-1$ (where we identify $-1$ with $3$) are
related to the gapless states at the edge $s$ via a $\pi/2$-rotation
(see SM~\cite{SM}). Hence, the expectation values of $S_{\perp}$ in
the basis of states $\Ket{\Psi^{s}_{0,\pm}}$ is exactly equal to the
expectation values of $S_{\parallel}$ in the basis
$\Ket{\Psi^{s-1}_{0,\pm}}$, resulting in
$\Braket{\Psi^{s}_{\pm} | S_{\perp} | \Psi^{s}_{\pm}} = \pm
\trm{sgn}\left( m_s \right) m_{s-1} / \Delta_{\trm{z}}$. Taking into
account that, restricted by the inversion symmetry, the sign of $m_s$
changes on every second edge, the sign of the perpendicular component
of the spin polarization of low-energy states changes on every edge. We refer to this
  feature as the quadrupolar structure of the spin polarization.

In Fig.~\ref{fig:soti_inversion_stm} we present the results of a
numerical calculation of the spin polarization, based on the Keldysh
formalism~\cite{MullerEtAl2020}. We mimic the STM
measurement~\cite{JozwiakEtAl2016, JeonEtAl2017, JackEtAl2019,
  KimEtAl2019, RubyAl2019} by coupling the system to a spin-polarized
tip with an amplitude $\kappa$. We then calculate the current $J$ that
flows between the system and the tip by summing up the contribution of
the states in the energy window $[-V, 0]$ (see SM~\cite{SM}), where
$V$ corresponds to the bias voltage of the STM tip. We consider two
directions of the tip polarization: one parallel and one perpendicular
to the Zeeman field. As expected from the analytical argument at
$k_s = 0$, we find that in the topological regime, the expectation value of
$S_{\parallel}$ is roughly constant and negative, while
for $S_{\perp}$ it changes sign on every edge. These features are characteristic for the SOTSCs close to the topological phase transition.

We note that the magnitude of spin polarization may depend on the properties of the states $\Ket{\Psi^{s}_{0,\pm}}$ and the symmetries of $\mc{H}_{0}$ (see SM~\cite{SM} for a more detailed study including the calculation of the quadrupolar moment across the phase transition). Nevertheless, the quadrupolar structure associated with the sign of the spin polarization at different edges depends only on the sign of $m_{s}$, which are topologically protected quantities. Hence, observing a quadrupolar structure of the spin-polarization provides a prominent and unique signature of the system topology. Moreover, the effective edge Hamiltonian $\mc{H}_{\trm{eff},s}$ as well as the argument justifying the emergence of MCSs are generally valid for
inversion-symmetric SOTSCs~\cite{ShiozakiSato_2014, GeierEtAl_2018,
  Khalaf_2018, Trifunovic_2020}.

\begin{figure}[t]
  \centering
  \includegraphics[width=.99\columnwidth]{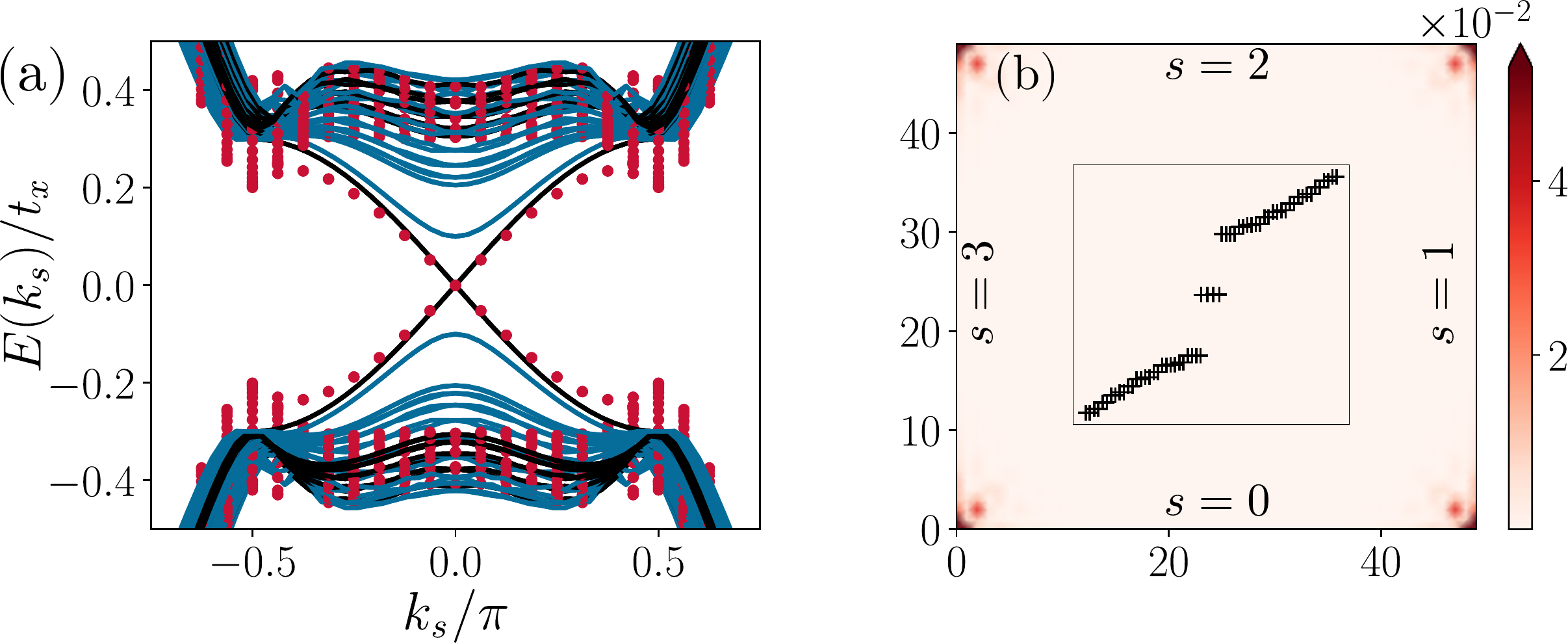}
  \caption{(a) Spectrum of $\mc{H}'$ for $\Delta_{\trm{sc}} = 0$. When
    $\Delta_{\trm{z}} = 0$ (black lines), the system hosts a pair of
    gapless helical edge modes. A finite
    value of $\Delta_{\trm{z}} = 0.1 t_x$ gaps out the edge modes in a
    geometry with OBC along the $y$ axis (blue lines, $k_s \equiv k_x$) but
    leaves the edge modes gapless in a geometry with OBC along the $x$ axis
    (red dots, $k_s \equiv k_y$). (b) Probability density of the MCSs
    calculated in a geometry with OBCs along both $x$ and $y$
    axes in a topological regime of the phase diagram with
    $\Delta_{\trm{z}} = 2 \Delta_{\trm{sc}} = 0.1 t_x$. The inset
    shows the low-energy spectrum. The remaining parameters are
    $\Gamma = t_y = t_x$, $\alpha_x = \alpha_y = 0.3 t_x$,
    $\mu_0 = 0$, and $a = 1$.}
  \label{fig:soti_mirror}
\end{figure}

{\it SOTSCs with four corner states.} Next, we generalize our findings
to a different class of SOTSCs and consider a SOTSC hosting a
quadruplet of MCSs~\cite{WangLinHughes_2018, LiuHeNori_2018,
  FrancaEfremovFulga_2019, Yan_2019, ZhangColeWeDasSarma_2019,
  Zhu_2019, LaubscherEtAl_2020, WuLiuThomaleLiu_2019, WuEtAl_2019,
  ZhangCalzonaTrauzettel_2020}. The basic ingredient of our
construction here is a minimalist version of a 2D
TI~\cite{TI_BernevigHughesZhang2006} proximity coupled to an $s$-wave
superconductor with amplitude $\Delta_{\trm{sc}}$. The
corresponding Hamiltonian reads
\begin{align}
  \label{eq:h0_mirror}
  & \mc{H}'_0({\vec{k}}) =
    \left\lbrace \Gamma -
    2 t_x \left[ 1 - \cos(k_x a) \right] -
    2 t_y \left[ 1 - \cos(k_y a) \right]
    \right\rbrace \eta_z \tau_z    
    \notag \\
  &
    - \mu_0 \eta_z +
    \left[ \alpha_x \sin(k_x a) \sigma_z \tau_x -
    \alpha_y \sin(k_y a) \eta_z \tau_y \right] +
    \Delta_{\trm{sc}} \eta_y \sigma_y ,
\end{align}
where Pauli matrices $\eta_j$, $\sigma_j$, and $\tau_j$ play exactly
the same role as in Eq.~\eqref{eq:h0_inversion}. The parameters $t_x$,
$t_y$, $\alpha_x$, $\alpha_y$, and $\Gamma$ depend on the microscopic
details of the system (see SM~\cite{SM}).
  For $\Delta_{\trm{sc}} = 0$ the
system is characterized by a topological phase transition as a
function of the parameter $\Gamma$. The closing of the gap occurs at
$\Gamma = 0$, independently of the value of $\alpha_x$ and
$\alpha_y$. The region $\Gamma < 0$ is topologically trivial, while
$\Gamma > 0$ is identified with a TI phase that supports gapless
helical edge modes protected by the time-reversal symmetry
$T = i \sigma_y K$. Again, we verify the presence of such edge modes
by analyzing the model numerically in a geometry with OBC along one
particular axis. The result of the calculation is shown in
Fig.~\ref{fig:soti_mirror}(a).

At non-zero $\Delta_{\trm{sc}}$, the helical edge modes are
gapped out. However, such a process acts identically on all the edges,
transforming the system into a trivial superconductor. In order to
achieve richer physics, one can apply the in-plane Zeeman
field, described by the Hamiltonian term
\begin{align}
  \label{eq:hz_mirror}
  \mc{H}'_{\trm{z}}
  = \Delta_{\trm{z}} \eta_z \sigma_x ,
\end{align}
such that the total Hamiltonian becomes
$\mc{H}'(\vec{k}) = \mc{H}'_0({\vec{k}}) + \mc{H}'_{\trm{z}}$. The
resulting system is invariant under the inversion symmetry
$I = \tau_z$. We also note that the exact orientation of the Zeeman
field in the $xy$ plane is not important, since the spectrum is
invariant under an arbitrary rotation around the spin quantization
axis. Interestingly, the effect of the in-plane Zeeman field alone
differs strongly depending on the edge considered. This can be seen by considering two special lines in the momentum space,
corresponding to vanishing $k_x$ or $k_y$, which can be used to
describe the physics of the system in a geometry with OBC along the
$y$ or $x$ axis, respectively (see SM~\cite{SM}). For $k_y = 0$, the
Zeeman term leaves the edge modes gapless, while for $k_x = 0$ it
leads to an opening of the gap of the size $\Delta_{\trm{z}}$. As a
consequence, the modes propagating along $x$-edges of the system are
gapped, while the ones along the $y$-edges remain gapless. The numerical
verification of this feature is shown in
Fig.~\ref{fig:soti_mirror}(a).

By taking into account the effect of both the superconducting and
Zeeman terms, one can construct the low-energy effective $4 \times 4$
edge Hamiltonian, which reads
\begin{align}
  \label{eq:heff_mirror}
  \mc{H}'_{\trm{eff}, s}(k_{s})
  = v_{\trm{F}} k_{s} \rho_z + m_{s} \eta_z \rho_x + \Delta_{\trm{sc}} \eta_y \rho_y.
\end{align}
Here, we use the same convention as in Eq.~(\ref{eq:heff_inversion})
and denote by $\rho_j$ the matrices acting on the states
$\Ket{\Psi'^{s}_{0,\pm}}$, associated with the right- and left-moving
modes living on the edge $s \in \left\lbrace 0, 1, 2, 3 \right\rbrace$
[see Fig.~\ref{fig:soti_mirror}(b)]. The matrices $\eta_j$ act in the
particle-hole space and $m_s$ denotes the mass originating from the
Zeeman term. The mass vanishes on two $y$-edges: $m_1 = m_3 = 0$,
while $|m_0| = |m_2| = \Delta_{\trm{z}}$ on two $x$-edges. Moreover,
the effective description of every edge is identical to the low-energy
physics of a topological nanowire~\cite{FuKane2009, NilssonEtAl2008,
  LutchynEtAl2010, OregEtAl2010, AliceaEtAl2010}, characterized by the
topological phase transition at a critical point
$\Delta_{\trm{z}} = \Delta_{\trm{sc}}$. Hence, if the Zeeman field is
strong enough such that $\Delta_{\trm{z}} > \Delta_{\trm{sc}}$, the
two $x$-edges of the system correspond to two wires in the topological
regime, while the two $y$-edges -- to two trivial nanowires. As a
result, four corners of the system host four zero energy states,
identified with MCSs. The regime
$\Delta_{\trm{z}} < \Delta_{\trm{sc}}$ is the topologically trivial
phase. Similarly, the corner states disappear when the system moves
into the trivial region of the phase diagram with $\Gamma < 0$. We
also note that the topological description of the present model does
not rely on the presence of the inversion symmetry $I$. Instead, the
existence of four MCSs is ensured by a particular spatial structure of
the gapping processes $m_s$ and $\Delta_{\trm{sc}}$, namely by the
fact that one pair of opposite edges has $|m_s| < \Delta_{\trm{sc}}$,
while the other one has $|m_s| > \Delta_{\trm{sc}}$.

\begin{figure}[t]
  \centering
  \includegraphics[width=.99\columnwidth]{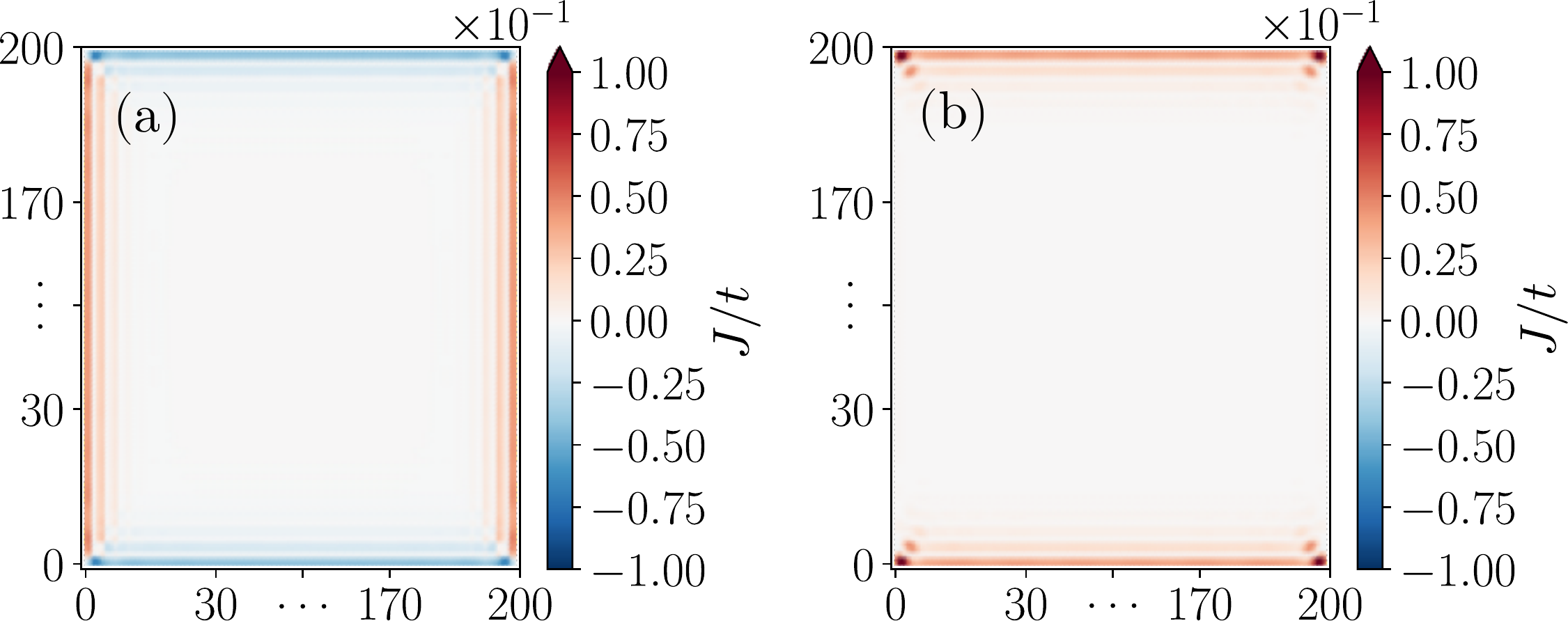}
  \caption{Numerical calculation of the current $J$ through the
    spin-polarized STM tip for a system described by $\mc{H}'$ given
    below Eq.~(\ref{eq:hz_mirror}). The polarization of the STM is
    chosen to be parallel to the Zeeman field. (a) In the topological
    phase with $\Delta_{\trm{z}} = 1.75 \Delta_{\trm{sc}}$, the
    polarization changes sign on every edge, resulting in a
    quadrupolar structure similar to the one of
    Fig.~\ref{fig:soti_inversion_stm}. (b) In the topologically
    trivial phase with $\Delta_{\trm{z}} = 0.25 \Delta_{\trm{sc}}$,
    the polarization has the same sign along the entire system
    boundary. The signal is weaker along the $y$-edges because less edge states contribute in the energy window. The remaining parameters of the simulations are $\Gamma = t_y = t_x$, $\alpha_y = \alpha_x = 0.3 t_x$
    $\Delta_{\trm{sc}} = 0.1 t_x$, $\kappa = 5 \cdot 10^{-4} t_x$. The
    current contributions are summed up over $V = \Delta_{\trm{sc}}$.}
  \label{fig:soti_mirror_stm}
\end{figure}

{\it Quadrupolar structure of $\Braket{S_{\parallel}}$.} We show that,
again, a spin-polarized STM~\cite{MullerEtAl2020, JozwiakEtAl2016, JeonEtAl2017, JackEtAl2019, KimEtAl2019, RubyAl2019} can be used
to probe the topological phase diagram of $\mc{H}'_{\trm{eff},s}$. To
show this, we consider the edges $s = 0, 2$ and focus on the physics
at the Dirac point $k_s = 0$, where the phase transition occurs. We
denote by $\Ket{\Psi'^{s}_{\pm}}$ two eigenstates of
$\mc{H}'_{\trm{eff},s}$ associated with two lowest magnitude
eigenvalues $\pm |m_s - \Delta_{\trm{sc}}|$. The states
$\Ket{\Psi'^{s}_{\pm}}$ are eigenstates of the Zeeman term and are
characterized by the polarization
$\Braket{\Psi'^{s}_{\pm} | S_{\parallel} | \Psi'^{s}_{\pm}} = \pm 1$
for $\Delta_{\trm{z}} > \Delta_{\trm{sc}}$. The polarization of these
two states remains constant up to the point
$\Delta_{\trm{z}} = \Delta_{\trm{sc}}$, at which the gap is closed at
the edges and the topological phase transition occurs. Decreasing the
Zeeman term further, the spin polarization of these two states flip,
acquiring a new value
$\Braket{\Psi'^{s}_{\pm} | S_{\parallel} | \Psi'^{s}_{\pm}} = \mp
1$. At the same time, no gap closing occurs at the edges $s=1,3$. As a
consequence, the expectation value of $S_{\parallel}$ grows smoothly
as a function of the ratio ratio
$\Delta_{\trm{z}} / \Delta_{\trm{sc}}$ without flipping its
sign. Hence, the parallel component of the spin polarization changes
sign only on two $m_s \neq 0$ edges of the system as one goes through
the topological phase transition and it acquires the quadrupolar
structure only in the topological phase but not in the trivial
one (see SM~\cite{SM}). The expectation values of  perpendicular components of the
spin polarization are zero. This feature allows one to
unambiguously identify the topological phase transition occurring in such a system.

These analytical predictions in low-energy approximation are confirmed by a numerical study of
$\mc{H}'$, presented in Fig.~\ref{fig:soti_mirror_stm}, where we
calculate the current through the spin-polarized STM tip (see
SM~\cite{SM}). When $\Gamma > 0$, we find that the expectation values
of $S_{\parallel}$ are always positive along the two $y$-edges and are
positive (negative) when $\Delta_{\trm{z}} < \Delta_{\trm{sc}}$
($\Delta_{\trm{z}} > \Delta_{\trm{sc}}$) along the two $x$-edges. When
$\Gamma < 0$, the description in terms of $\mc{H}_{\trm{eff},s}$
breaks down and the bulk signal dominates over the edge signal. As
expected, the perpendicular components of the spin polarization is
found to be negligibly small everywhere except at the system
corners. We have also checked that the proposed feature is stable
against weak disorder.

{\it Conclusions.}  In this work, we analyzed the spin polarization of
two-dimensional topological superconductors as a signature indicating
topological phases. In particular, we showed that a spin-polarized STM
can be used to determine the topological phase diagram in two types of
second-order topological superconductors (SOTSCs) with broken
time-reversal symmetry. In SOTSCs, which host a pair of corner states,
the distinguishing feature of the topological phase is the quadrupolar
structure of the spin polarization perpendicular to the Zeeman
field. Similarly, in SOTSCs with two pairs of corner states, the spin
polarization parallel to the Zeeman field acquires such a quadrupolar
structure only in the topological phase. This probe can be used in
conjunction with the usual experimental protocols, such as the state
tomography and the measurement of the differential conductance, to
verify the topological nature of bound states and serves as an
additional independent signature of the topological phase transition.

\section*{Acknowledgments}

We gratefully acknowledge many useful discussions with Katharina
Laubscher, Flavio Ronetti, and Ferdinand Schulz. This work was
supported by the Swiss National Science Foundation, NCCR QSIT, and the
Georg H. Endress foundation as well as the Deutsche
Forschungsgemeinschaft via RTG 1995 and by the Deutsche
Forschungsgemeinschaft (DFG, German Research Foundation) under
Germany's Excellence Strategy - Cluster of Excellence Matter and Light
for Quantum Computing (ML4Q) EXC 2004/1 - 390534769. This project
received funding from the European Union's Horizon 2020 research and
innovation program (ERC Starting Grant, grant agreement No 757725). We
acknowledge support from the Max Planck-New York City Center for
Non-Equilibrium Quantum Phenomena. Simulations were performed with
computing resources granted by RWTH Aachen University under projects
rwth0498, rwth0563, rwth0564 and rwth0593.


\bibliographystyle{unsrt}


\clearpage
\onecolumngrid
\begin{center}
   \textbf{\large Supplemental Material: Quadrupole spin polarization as  signature of  second-order  topological superconductors}\\
  \vspace{8pt}
  Kirill Plekhanov$^{1}$, Niclas M\"{u}ller$^{2}$, Yanick Volpez$^{1}$, Dante M. Kennes$^{2}$, Herbert Schoeller$^{2}$, Daniel Loss$^{1}$, and Jelena Klinovaja$^{1}$
  \\ \vspace{4pt}
  $^{1}$ {\it Department of Physics, University of Basel, Klingelbergstrasse 82, CH-4056 Basel, Switzerland} \\ \vspace{4pt}
  $^{2}$ {\it Institut f\"{u}r Theorie der Statistischen Physik, RWTH Aachen, 52056 Aachen, Germany and JARA - Fundamentals of Future Information Technology}
\end{center}
\twocolumngrid

\setcounter{section}{0}
\setcounter{equation}{0}
\setcounter{figure}{0}
\setcounter{page}{1}
\makeatletter
\renewcommand{\thesection}{S\arabic{section}}
\renewcommand{\theequation}{S\arabic{equation}}
\renewcommand{\thefigure}{S\arabic{figure}}
\titleformat{\section}[hang]{\large\bfseries}{\thesection.}{5pt}{}

\section{\label{sec:model_details} Microscopic details of the models}

In this section, we provide microscopic details on the Hamiltonians $\mc{H}_0$ and $\mc{H}'_0$ considered in the main text of the manuscript. We also describe possible venues for an experimental realization of corresponding physical systems.

Firstly, we consider the Hamiltonian
\begin{align}
  & \mc{H}_0({\vec{k}}) = 2t
    \left[ 2 - \cos(k_x a) - \cos(k_y a) \right] \eta_z - \mu_0 \eta_z +
    \Gamma \eta_z \tau_x
    \notag \\
  & +
    \alpha \left[ \sin(k_y a) \sigma_x - \sin(k_x a) \eta_z \sigma_y \right] \tau_z +
    \Delta_{\trm{sc}} \eta_y \sigma_y \tau_z
\end{align}
from Eq.~\eqref{eq:h0_inversion} of the main text. This Hamiltonian
was first introduced in Ref.~\cite{VopezEtAl_2019}, where it was used
to describe a system composed of two Rashba layers that are tunnel
coupled to each other with an amplitude $\Gamma$. Here, $t$ is the
term proportional to the kinetic energy of the electrons, which is the
same in both layers. Moreover, each layer is assumed to have Rashba
spin-orbit interactions with an amplitude $\alpha$ of opposite
signs. In what follows, we consider $t, \alpha > 0$. The chemical potential
$\mu_0$ is tuned to the spin-orbit crossing point at $\vec{k} =
0$. Finally, the system is coupled to a 2D $s$-wave superconducting
Josephson junction with a phase factor of $\pi$, inducing a
superconducting pairing $\Delta_{\trm{sc}}$ via the proximity
effect. The Pauli matrices $\eta_j$ act on the particle-hole space,
$\sigma_j$ -- on the spin space, and $\tau_j$ -- on the space
associated with two Rashba layers. The lattice spacing is denoted
  by $a$.

Secondly, we consider the Hamiltonian
\begin{align}
  & \mc{H}'_0({\vec{k}}) =
    \left\lbrace \Gamma -
    2 t_x \left[ 1 - \cos(k_x a) \right] -
    2 t_y \left[ 1 - \cos(k_y a) \right]
    \right\rbrace \eta_z \tau_z    
    \notag \\
  &
    - \mu_0 \eta_z +
    \left[ \alpha_x \sin(k_x a) \sigma_z \tau_x -
    \alpha_y \sin(k_y a) \eta_z \tau_y \right] +
    \Delta_{\trm{sc}} \eta_y \sigma_y
\end{align}
from Eq.~\eqref{eq:h0_mirror} of the main text. In the regime when
$t_x \sim t_y$ and $\alpha_x \sim \alpha_y$, $\mc{H}'_{0}$ can be seen
as a modified version of the BHZ
Hamiltonian~\cite{TI_BernevigHughesZhang2006}, where we neglect
the usual kinetic term (which does not have any effect on the
topological description), but where we take into account the anisotropy effect
in the $xy$-plane. In this case, the degrees of freedom associated
with $\tau_j$ correspond to electron/hole orbitals. The
hopping amplitudes $t_x$ and $t_y$ describe antisymmetric components
of the kinetic term, while $\alpha_x$ and $\alpha_y$ are the
spin-orbit interaction amplitudes. In what follows, all these parameters are
  assumed to be strictly greater than zero. The parameter $\Gamma$ is
responsible for the topological phase transition between a trivial
insulator and the TI. In a quantum well experimental setup, $\Gamma$
depends on the thickness of the quantum well~\cite{koenig2007,
  molenkamp2009, wangzhang2008}.

In addition to the BHZ model, $\mc{H}'_{0}$ can also be generated
using a coupled wire construction~\cite{KaneEtAl2002, TeoAndKane2014}
in a strongly anisotropic regime with $t_y \ll t_x$,
$\alpha_y \ll \alpha_x$. In this case, $t_x$ and $\alpha_x$ correspond
to the kinetic and Rashba terms along the wire direction, while $t_y$
and $\alpha_y$ correspond to the inter-wire couplings, which are much
smaller in amplitude.

\section{\label{sec:stm} Numerical methods}

In an STM measurement, the tip of the microscope, biased at the
voltage $V$, is brought close to the sample. This results in  a current flowing 
 through the tip into the sample, the amplitude of
which depends on the LDOS of the sample and on the overlap between the
wavefunctions of the sample and the tip. Hence, if the tip is
spin-polarized, the STM measurement will provide additional
information on the spin polarization of the sample. Moreover, the
voltage difference between the sample and the tip determines how many
eigenstates contribute to the current. We note that, alternatively,
one can use spin-polarized quantum dots \cite{denis,denis2,tip2},
however, they are less mobile.

\begin{figure}[t]
  \centering
  \includegraphics[width=.99\columnwidth]{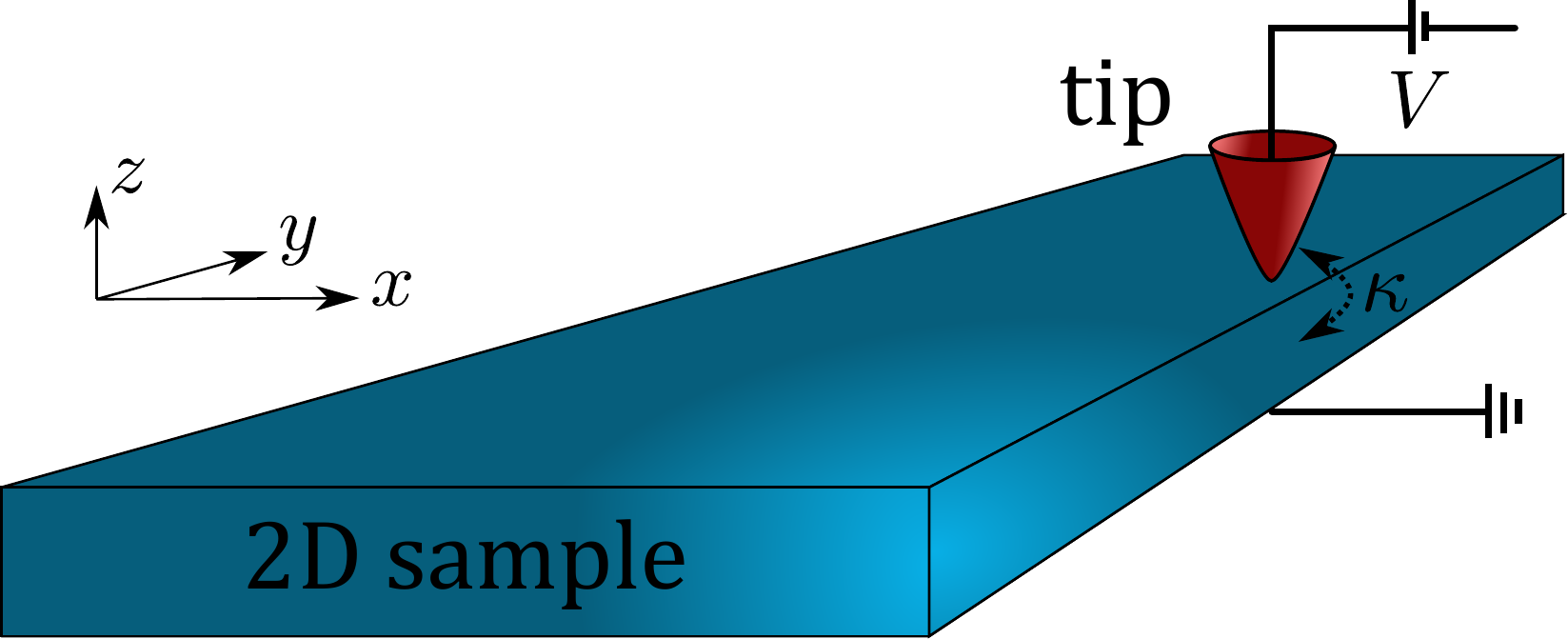}
  \caption{Schematic representation of the setup. The spin-polarized STM tip biased at voltage $V$ is brought close to the 2D sample which realizes a SOTSC. The current measured using STM allows one to probe the topological phase diagram of the SOTSC.}
  \label{fig:setup}
\end{figure}

In our work, we model the STM measurement and perform numerical simulations
using the Keldysh formalism~\cite{MullerEtAl2020}. To do this calculation, we consider
the following setup consisting of two parts: the sample, corresponding
to a 2D SOTSC, and the STM tip, are described by the Hamiltonians
$\mc{H}_{\trm{sam}}$ and $\mc{H}_{\trm{tip}}$, respectively. The STM
tip is tunnel coupled to the sample with an amplitude denoted by
$\kappa$ (see Fig.~\ref{fig:setup}). The sample is simulated on a
square lattice, and the tunneling between the tip and the sample is
assumed to occur locally at one given site of the lattice, depending
on the position of the tip. The effect of the applied voltage $V$
induces the difference of the chemical potential $e V$ between the tip
and the sample, with $e$ -- the electron charge. The influence of the
tip is encoded in the induced self-energy which dresses the bare
Green's function of the sample. The resulting time-dependent current flowing from the probe reservoir into the sample at time $t$ can be expressed
using the Keldysh formalism as
\begin{equation}
    J(t) = -e\int \trm{d}t^{\prime} \trm{Tr} \left\{\Sigma^R(t,t^{\prime}) G^<(t^{\prime},t) - G^<(t,t^{\prime}) \Sigma^A(t^{\prime},t)\right\},
\end{equation}
where $\Sigma^{\trm{R}/\trm{A}}$ are the retarded/advanced
self-energies of the STM tip, $G^{\lessgtr}$ are the fully dressed
greater/lesser Green's functions of the sample, and $e$ is the
  electron charge which is assumed to be equal to one. In the wide
band limit and at the equilibrium, the above expression simplifies to
\begin{equation}
  \label{eq:current_sup}
    J = e\trm{Tr} \int \frac{\trm{d}{\omega}}{2\pi}\Sigma^<(\omega) G^R(\omega) \Sigma^<(\omega) G^A(\omega).
\end{equation}
In the main part of the manuscript, we mostly focus on the features of
the sample close to the boundary, and show that they provide an
alternative way to probe the topological phase transition in some
classes of SOTSCs. We also present how the result of such a
calculation varies as a function of the tip polarization and the
sample Hamiltonian $\mc{H}_{\trm{sam}}$, describing different types of
SOTSCs. In the section``Quadrupolar structure of $\Braket{S_{\perp}}$" the
sample Hamiltonian is taken to be $\mc{H}$, while in the section
``Quadrupolar structure of $\Braket{S_{\parallel}}$" it is $\mc{H}'$.

\section{Analytical calculation of  edge states and  spin polarization}

In this section, we provide details on the analytical calculation of
the wavefunctions associated with helical states in both models
considered in the main text. We also study the effect of the
perturbations, which are added to the models in order to gap out the
helical modes, and calculate the expectation values of the spin
operators to deduce their spin polarization.

\subsection{SOTSC with two corner states \label{sec:appendix_edgeStates_inversion}}

\subsubsection{Properties of the $s = 0$ edge}

We start with the model described by the Hamiltonian $\mc{H}_{0}$ from
Eq.~\eqref{eq:h0_inversion}~\cite{VopezEtAl_2019}. We fix the
  chemical potential as $\mu_0 = 0$ and take the lattice spacing $a$
  to be equal to one.  First, we consider the $s=0$ edge localized at
$y=0$. We focus on the physics at the point $k_x = 0$, which describes
the solutions uniform along the $x$ axis. We also linearize the
  resulting problem around the Fermi momenta
  $k_{\trm{F}, \trm{i}} = 0$ and
  $k_{\trm{F}, \trm{e}} = \trm{arctan}(\alpha / t)$~\cite{lin}. In
  order to do this, we switch to the basis of slowly varying left and
  right movers
\begin{align}
  \label{eq:movers}
  \tilde{\psi}_{\uparrow 1}
  &=
    \tilde{L}_{\uparrow 1} e^{-i k_{\trm{F}, \trm{e}} y} +
    \tilde{R}_{1 \uparrow}, \quad
    \tilde{\psi}_{\downarrow 1}
    =
    \tilde{L}_{\downarrow 1} +
    \tilde{R}_{\downarrow 1} e^{i k_{\trm{F}, \trm{e}} y}, \quad
    \notag \\
  \tilde{\psi}_{\uparrow \bar{1}}
  &=
    \tilde{L}_{\uparrow \bar{1}} +
    \tilde{R}_{\uparrow \bar{1}} e^{i k_{\trm{F}, \trm{e}} y}, \quad
    \tilde{\psi}_{\bar{1} \downarrow}
    =
    \tilde{L}_{\bar{1} \downarrow} e^{-i k_{\trm{F}, \trm{e}} y} +
    \tilde{R}_{\bar{1} \downarrow}, \quad
\end{align}
defined such that $\tilde{\sigma} = \uparrow, \downarrow$ denotes the
spin projection onto the $x$ axis. In this new basis, the linearized
Hamiltonian reduces to
\begin{align}
  \tilde{\mc{H}}_{0}
  &=
    \alpha k_y \delta_z +
    \Gamma \eta_z
    \left( \tau_x \delta_x - \tau_y \delta_y \tilde{\sigma}_z \right) \eta_z / 2 +
    \Delta_{\trm{sc}} \eta_y \tilde{\sigma}_y \tau_z \delta_x ,
\end{align}
where the Pauli matrices $\delta_j$ act on the space of left and right movers. The
obtained problem can be solved by substituting $k_y = -i \partial_y$
and imposing  vanishing boundary condition at $y = 0$. We find that
in the region $\Gamma > \Delta_{\trm{sc}}$, the system admits two
zero-energy solutions described by the following wavefunctions:
\begin{align}
  \label{eq:psi0_inversion}
  \Psi^0_{0,+}(y)
  &= \frac{1}{\mc{N}}
    \left( f_1, g_1, f_1^*, g_1^*, f_{\bar{1}}, g_{\bar{1}},
    f^*_{\bar{1}}, g^*_{\bar{1}} \right)^{T} ,
    \notag \\
  \Psi^0_{0,-}(y)
  &= \frac{1}{\mc{N}}
    \left( g^{*}_1, -f^{*}_1, g_1, -f_1, g^{*}_{\bar{1}}, -f^{*}_{\bar{1}},
    g_{\bar{1}}, -f_{\bar{1}} \right)^{T} ,
\end{align}
where we used Eq.~\eqref{eq:movers} to go back to the original basis
$(\psi_{\uparrow 1}, \psi_{\downarrow 1}, \psi^{\dag}_{\uparrow 1},
\psi^{\dag}_{\downarrow 1}, \psi_{\uparrow \bar{1}}, \psi_{\downarrow
  \bar{1}}, \psi^{\dag}_{\uparrow \bar{1}}, \psi^{\dag}_{\downarrow
  \bar{1}})$. Here $\mc{N}$ is the normalization constant, and the
functions $f_{\tau}$ and $g_{\tau}$ are expressed as
\begin{align}
  f_1 = g_{\bar{1}} = -i f^{*}_{\bar{1}} = - i g^{*}_{1} =
  e^{-y / \xi_{\trm{i}}} - e^{i k_{\trm{F}, \trm{e}} y} e^{-y / \xi_{\trm{e}}} ,
\end{align}
with $\xi_{\trm{i}} = \alpha / (\Gamma - \Delta_{\trm{sc}})$ and
$\xi_{\trm{e}} = \alpha / \Delta_{\trm{sc}}$. As expected from the
Kramers partner of gapless helical modes in a helical topological
superconductor, the two obtained solutions satisfy
$T \Ket{\Psi^0_{0,\pm}} = \pm \Ket{\Psi^0_{0,\mp}}$ and
$P \Ket{\Psi^0_{0,\pm}} = \Ket{\Psi^0_{0,\pm}}$, where $P = \eta_x K$
is the particle-hole symmetry operator.

By using the particular form of the solutions at $k_x = 0$, obtained
in Eq.~\eqref{eq:psi0_inversion}, we can now include perturbatively
the omitted Zeeman term as well as the first order kinetic term in
$k_x$. To begin with, we express the  term linear in $k_x$ as
\begin{align}
  \Braket{\Psi^0_{0,\pm} | -\alpha k_x \eta_z \sigma_y \tau_z | \Psi^0_{0,\pm}} &=
    \pm v_{\trm{F}} k_x ,
    \notag \\
  \Braket{\Psi^0_{0,+} | -\alpha k_x \eta_z \sigma_y \tau_z | \Psi^0_{0,-}} &=
    0 ,
\end{align}
where $v_{\trm{F}} = \alpha \Delta / \Gamma$ is the Fermi velocity of
the edge modes. Similarly, the two components of the Zeeman term can
be expressed as
\begin{align}
  \label{eq:gaps_inversion}
&  \Braket{\Psi^0_{0,\pm} |
  \Delta_{\trm{z}} \cos(\theta_{\trm{z}})\eta_z \sigma_x
  | \Psi^0_{0,\pm}} =
    0 ,
    \notag \\
&  \Braket{\Psi^0_{0,+} |
  \Delta_{\trm{z}} \cos(\theta_{\trm{z}})\eta_z \sigma_x
  | \Psi^0_{0,-}} =
    i \Delta_{\trm{z}} \cos(\theta_{\trm{z}}) ,
    \notag \\
&  \Braket{\Psi^0_{0,\pm} |
    \Delta_{\trm{z}} \sin(\theta_{\trm{z}})\sigma_y
  | \Psi^0_{0,\pm}} =
    0 ,
    \notag \\
&  \Braket{\Psi^0_{0,+} |
  \Delta_{\trm{z}} \sin(\theta_{\trm{z}})\sigma_y
  | \Psi^0_{0,-}} =
    0 .
\end{align}
Combining these two results, we recover the effective Jackiw-Rebbi Hamiltonian
\begin{align}
  \mc{H}_{\trm{eff}, 0}(k_{x})
  = v_{\trm{F}} k_{x} \rho_z - \Delta_{\trm{z}} \cos(\theta_{\trm{z}}) \rho_y ,
\end{align}
which describes the low-energy physics of the edge $s = 0$.

\subsubsection{Remaining edges and quadrupolar structure of the spin polarization}

Next, we obtain similar results for the three remaining edges of the system by using the rotational symmetry of the Hamiltonian $\mc{H}_{0}$. The rotational symmetry operator can be explicitly written down as $U_{\trm{rot}}(\theta) = \trm{exp} \left[ i \theta \left( L_{z} + S_{z} \right) \right]$, where
$L_{z} = -i (x \partial_y - y \partial_x)$ and $S_{z} = \eta_z \sigma_z / 2$ are the $z$ component of the orbital momentum and the spin, respectively. The non-perturbed Hamiltonian satisfies $U_{\trm{rot}}(\theta) \mc{H}_0({\vec{k}}) U^{-1}_{\trm{rot}}(\theta) = \mc{H}_0({\vec{k}})$. As a consequence, the expression of the states at the three remaining edges can be obtained by using
$\Ket{\Psi^s_{0,\pm}} = U_{\trm{rot}}(\theta_s) \Ket{\Psi^0_{0,\pm}}$, where
$\theta_s = s \pi / 2$. The Zeeman term is not invariant under the rotation symmetry transformation:
\begin{align} 
  U^{-1}_{\trm{rot}}(\theta) \mc{H}_{\trm{z}} U_{\trm{rot}}(\theta) = \Delta_{\trm{z}} \left[
  \cos(\theta - \theta_{\trm{z}})\eta_z \sigma_x +
  \sin(\theta - \theta_{\trm{z}})\sigma_y
  \right] .
\end{align}
Combining this with the results of Eq.~\eqref{eq:gaps_inversion}, we
deduce that the gap opened by the Zeeman term on the edge $s$ can be
expressed as
$m_s = \Delta_{\trm{z}} \cos(\theta_{\trm{z}} - \theta_s)$. We note
that this gap, indeed, satisfies $m_0 = -m_2$ and $m_1 = -m_3$, as
required by the inversion symmetry.

Finally, we calculate the expectation values of the spin operators
$S_{\parallel} = \cos(\theta_z) \eta_z \sigma_x +
\sin(\theta_{\trm{z}}) \sigma_y$ and
$S_{\perp} = \cos(\theta_{\trm{z}}) \sigma_y - \sin(\theta_z) \eta_z
\sigma_x$ in the basis of the eigenstates of the effective Hamiltonian
$\mc{H}_{\trm{eff},s}$ at $k_s = 0$. These states diagonalize
$\mc{H}_{\trm{z}}$ and, hence, can be found as
$\Ket{\Phi^{s}_{\pm}} = (\Ket{\Psi^{s}_{0,+}} \mp i
\Ket{\Psi^{s}_{0,-}}) / \sqrt{2}$, in correspondence to the
eigenvalues $\pm m_s$. By ordering these states according to their
eigenvalues, we recover the two states $\Ket{\Psi^{s}_{\pm}}$ of the
main text corresponding to the eigenvalues $\pm  |m_s|$. Trivially, the parallel component of the spin polarization
of these states is equal to
\begin{align}
  \Braket{\Psi^{s}_{\pm} | S_{\parallel} | \Psi^{s}_{\pm}} =
  \pm |m_s| / \Delta_{\trm{z}} = \pm |\cos(\theta_{\trm{z}} - \theta_s)|.
\end{align}
In order to calculate the perpendicular in-plane component of the
polarization, we make use of the fact that
\begin{align}
  & \Braket{\Phi^{s}_{\pm} | S_{\perp} | \Phi^{s}_{\pm}}
 = 
    \Braket{\Phi^{s}_{\pm} | U_{\trm{rot}}(\pi/2) S_{\parallel} U^{-1}_{\trm{rot}}(\pi/2) | \Phi^{s}_{\pm}}
    \notag \\
  &\hspace{20pt}= 
    \Braket{\Phi^{s-1}_{\pm} | S_{\parallel} | \Phi^{s-1}_{\pm}} = \pm
    m_{s-1} / \Delta_{\trm{z}}.
\end{align}
Hence, once the states on the $s$-edge  are ordered, we obtain
\begin{align}
  \Braket{\Psi^{s}_{\pm} | S_{\perp} | \Psi^{s}_{\pm}}
  & =
  \pm \trm{sgn}(m_s) m_{s-1} / \Delta_{\trm{z}}
  \notag \\
  & =  \pm \trm{sgn}\left( m_s \right) \cos (\theta_{\trm{z}} - \theta_{s-1}).
\end{align}

\begin{figure}[t]
  \centering
  \includegraphics[width=.99\columnwidth]{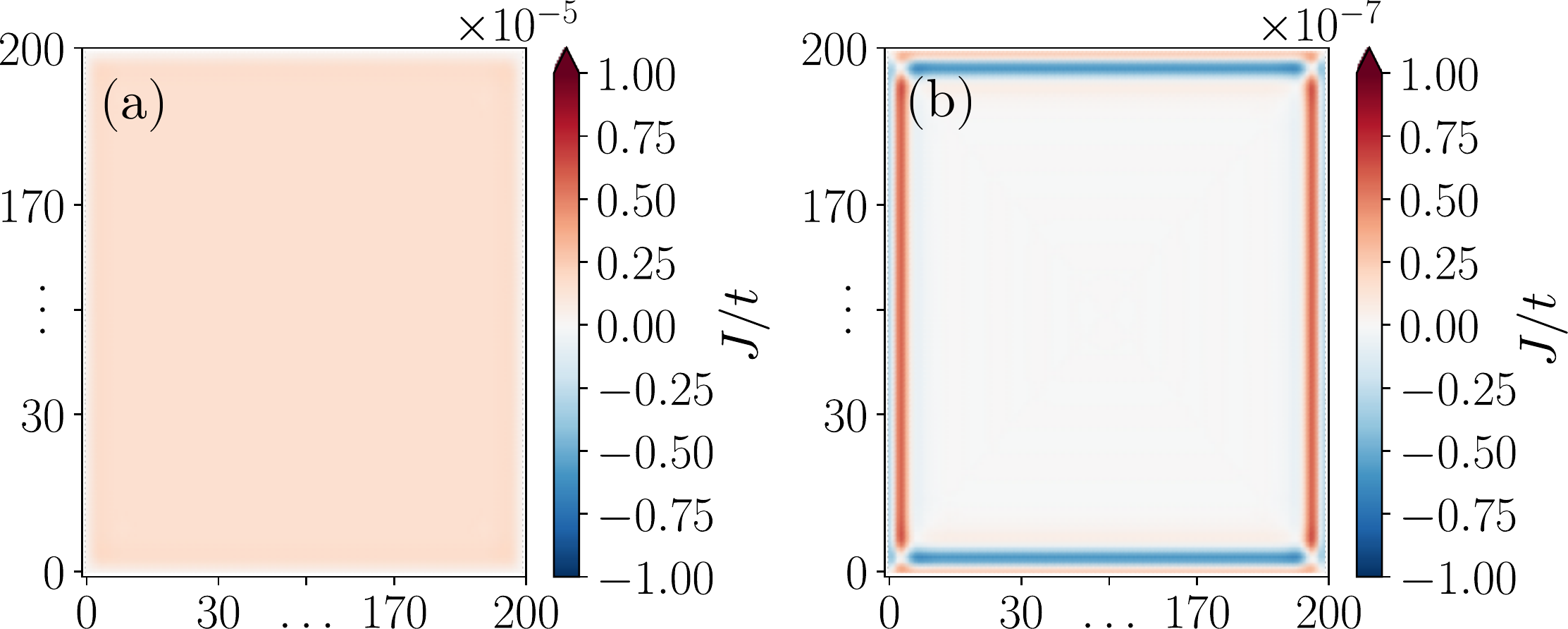}
  \caption{Numerical calculation of the current $J$ through the
    spin-polarized STM tip in the topologically trivial phase of
    $\mc{H}$ with $\Gamma = 0.25 \Delta_{\trm{sc}} = 0.06 t$. The
    polarization of the STM is chosen to be (a) parallel to the Zeeman
    field and (b) perpendicular to the Zeeman field. The remaining
    parameters are the same as in Fig.~\ref{fig:soti_inversion_stm}.
    The quadrupolar structure of $J$ is absent for the parallel
    component of the spin polarization. We observe the reminiscence of
      the quadrupolar structure for the perpendicular component,
      however the resulting current is several orders of magnitude
      smaller than in the topological phase (see
      Fig.~\ref{fig:soti_inversion_stm})}.
  \label{fig:soti_inversion_stm_trivial}
\end{figure}

These theoretical arguments are confirmed by numerical simulations of
a current flowing through the spin-polarized tip in an STM measurement
setup, see Fig.~\ref{fig:soti_inversion_stm} for a topological phase
with $\Gamma = 1.75 \Delta_{\trm{sc}}$ and
Fig.~\ref{fig:soti_inversion_stm_trivial} for a topologically trivial
phase with $\Gamma = 0.25 \Delta_{\trm{sc}}$. We find that, in the
topological phase, the current is strongest at the edges. When the tip
polarization is parallel to the direction of the Zeeman field, the
current is approximately uniform and negative along the entire
boundary of the systems. However, when the tip polarization is
perpendicular to the Zeeman field (but still lying in the $xy$ plane),
the current acquires a quadrupolar structure. In the trivial phase,
only bulk states contribute to the current
and the resulting signal is a few orders of magnitude smaller than in
the topological phase.

\subsection{SOTSC with four corner states \label{sec:appendix_edgeStates_mirror}}

\subsubsection{Properties of the $s = 0$ edge}

In the second part of this section, we study the low-energy physics of
the model described by the Hamiltonian $\mc{H}'_{0}(\bf k)$, see
Eq.~\eqref{eq:h0_mirror}. Similarly to
Section~\ref{sec:appendix_edgeStates_inversion}, we start by
considering the $s=0$ edge at $k_x = 0$, corresponding to a uniform
solution localized at $y = 0$. We fix $\mu_0 = 0$ and
set  $a=1$. Moreover, for a moment, we neglect the superconducting
pairing term $\Delta_{\trm{sc}}$, which we will include later
perturbatively. This allows us to focus only on electron or hole parts
of the spectrum.  To describe the electron part of the spectrum, we
choose the basis
$(\psi_{\uparrow 1}, \psi_{\downarrow 1}, \psi_{\uparrow \bar{1}},
\psi_{\downarrow \bar{1}})$. We expand the Hamiltonian up to 
second order in $k_y$ around $k_y=0$, which leads us to
\begin{align}
  \label{eq:ham0_solution_inversion_1}
  \mc{H}'_{0}(k_x=0, k_y) \approx
  \left( \Gamma - t_y k_y^2 \right) \tau_z - \alpha_y k_y \tau_y .
\end{align}
We proceed by substituting $k_y = -i \partial_y$ and looking for zero-energy solutions of the resulting equation. After imposing vanishing boundary condition at $y = 0$, we find two exponentially decaying solutions of the form
\begin{align}
  \label{eq:psi0_inversion_1}
  \Psi'^0_{0,+}(y)
  &=
    \frac{e^{-y / \xi_+} - e^{-y / \xi_-}}{\mc{N}}
    \left[ 1, 0, 1, 0 \right]^{T} ,
    \notag \\
  \Psi'^0_{0,-}(y)
  &=
    \frac{e^{-y / \xi_+} - e^{-y / \xi_-}}{\mc{N}} 
    \left[ 0, 1, 0, 1 \right]^{T} ,
\end{align}
where $\mc{N}$ is the normalization constant. Such solutions exist
only for $\Gamma > 0$, and the parameters $\xi_{\pm}$ are given by
\begin{align}
 \xi^{-1}_{\pm} = \frac{1}{2 t_y}
  \left( \alpha_y \pm \sqrt{\alpha_y^2 - 4 \Gamma t_y}\, \right) .
\end{align}
The two states are related by the time-reversal symmetry such that
$T \Ket{\Psi'^0_{0,\pm}} = \pm \Ket{\Psi'^0_{0,\mp}}$. We note that
the Hamiltonian defined in Eq.~\eqref{eq:ham0_solution_inversion_1}
does not contain Pauli matrices describing the spin space. Hence, the
spin quantization axis of the states $\Ket{\Psi'^0_{0,\pm}}$ could be
also chosen arbitrarily. Here, we decided to choose the form of the two
states as in Eq.~\eqref{eq:psi0_inversion_1}, such that it agrees with
the choice of the spin quantization axis from the main text and the
notation for the edge $s=1$.

Using the expression of the two states at $k_x = 0$, we calculate the expectation values of the Zeeman term and the kinetic term linear  in $k_x$. For the latter term, we obtain
\begin{align}
  \Braket{\Psi'^0_{0,\pm} | \alpha_x k_x \sigma_z \tau_x | \Psi'^0_{0,\pm}}
  &= \pm v_{\trm{F}} k_x ,
    \notag \\
  \Braket{\Psi'^0_{0,+} | \alpha_x k_x \sigma_z \tau_x | \Psi'^0_{0,-}}
  &= 0 ,
\end{align}
where $v_{\trm{F}} = \alpha_x$. The expectation values of the Zeeman
term are
\begin{align}
  \Braket{\Psi'^0_{0,\pm} |
  \Delta_{\trm{z}} \sigma_x
  | \Psi'^0_{0,\pm}} &=
    0 ,
    \notag \\
  \Braket{\Psi'^0_{0,+} |
  \Delta_{\trm{z}} \sigma_x
  | \Psi'^0_{0,-}} &=
    \Delta_{\trm{z}} .
\end{align}

In order to take into account the  superconducting $s$-wave pairing, we introduce a pair of states $\Ket{\Psi'^{0*}_{0,\pm}}$, which correspond to the particle-hole partners of the states $\Ket{\Psi'^{0}_{0,\pm}}$. These states belong to the second block of the Nambu space. This allows us to write the effective low-energy Hamiltonian describing the system boundary as
\begin{align}
  \mc{H}'_{\trm{eff}, 0}(k_{x})
  = v_{\trm{F}} k_{x} \rho_z + \Delta_{\trm{z}} \eta_z \rho_x + \Delta_{\trm{sc}} \eta_y \rho_y .
\end{align}
Here, $\eta_j$ acts in the particle-hole space and $\rho_j$ -- in the space of the two edge states. We note that we recover Eq.~\eqref{eq:heff_mirror} from the main text with $m_0 = \Delta_{\trm{z}}$. Next, we focus again on the physics at  momentum $k_x = 0$ and diagonalize $\mc{H}'_{\trm{eff}, 0}$.  It is easy to see that the two eigenstates corresponding to the two lowest magnitude eigenvalues $\pm (\Delta_{\trm{z}} - \Delta_{\trm{sc}})$ can be expressed as
\begin{align}
    \Ket{\Phi'^{0}_{+}}
    &=
    \frac{1}{2}
    \left( 
      \Ket{\Psi'^{0}_{0,+}} + \Ket{\Psi'^{0}_{0,-}} -
      \Ket{\Psi'^{0*}_{0,+}} + \Ket{\Psi'^{0*}_{0,-}} \right) ,
    \notag \\
    \Ket{\Phi'^{0}_{-}}
    &=
    \frac{1}{2}
    \left( 
      \Ket{\Psi'^{0}_{0,+}} - \Ket{\Psi'^{0}_{0,-}} -
      \Ket{\Psi'^{0*}_{0,+}} - \Ket{\Psi'^{0*}_{0,-}} \right) .
\end{align}
From this, we deduce that
\begin{align}
  \Braket{\Phi'^0_{\pm} | S_{\parallel} | \Phi'^0_{\pm}} = \pm 1 , \quad
  \Braket{\Phi'^0_{\pm} | S_{\perp} | \Phi'^0_{\pm}} = 0 ,
\end{align}
where $S_{\parallel} = \eta_z \sigma_x$ and $S_{\perp} = \sigma_y$ are
the parallel and perpendicular components of the spin polarization,
respectively. Finally, we sort the states according to their energies,
from negative to positive, to arrive at
$ \Ket{ \Psi'^0_{\pm}}$. The initial ordering of the states
$\Ket{\Phi'^0_{\pm}}$ is correct in the regime
$\Delta_{\trm{z}} > \Delta_{\trm{sc}}$, but has to be changed when
$\Delta_{\trm{z}} < \Delta_{\trm{sc}}$. Hence, we deduce that in the
basis of new, correctly ordered states, the expectation values of the
spin polarization become equal to
$\Braket{\Psi'^0_{\pm} | S_{\parallel} | \Psi'^0_{\pm}} = \pm 1$ when
$\Delta_{\trm{z}} > \Delta_{\trm{sc}}$ and
$\Braket{\Psi'^0_{\pm} | S_{\parallel} | \Psi'^0_{\pm}} = \mp 1$
otherwise.

\subsubsection{Properties of the $s = 1$ edge}

In the same way as for the edge $s=0$, we do the calculations on the
second non-equivalent edge of the system denoted by $s = 1$ by
considering the physics at $k_y = 0$. After expanding the Hamiltonian
$\mc{H}'_{0}(\bf k)$ [see Eq.~\eqref{eq:h0_mirror}] to  second
order in $k_x$, we obtain
\begin{align}
  \label{eq:ham0_solution_inversion_2}
  \mc{H}'_{0}(k_x, k_y=0) \approx
  \left( \Gamma - t_x k_x^2 \right) \tau_z +
  \alpha_x k_x \sigma_z \tau_x .
\end{align}
By substituting $k_x = -i \partial_x$ and imposing  vanishing
boundary condition at $x = 0$, we find two exponentially decaying
solutions at zero energy which have the form
\begin{align}
  \label{eq:psi0_inversion_2}
  \Psi'^1_{0,+}(x)
  &=
    \frac{e^{-x / \xi_+} - e^{-x / \xi_-}}{\mc{N}}
    \left[ 1, 0, i, 0 \right]^{T} ,
    \notag \\
  \Psi'^1_{0,-}(x)
  &=
    \frac{e^{-x / \xi_+} - e^{-x / \xi_-}}{\mc{N}} 
    \left[ 0, 1, 0, -i \right]^{T} ,
\end{align}
where the parameters $\xi_{\pm}$ are given by
\begin{align}
  \xi^{-1}_{\pm} = \frac{1}{2 t_x}
  \left( \alpha_x \pm \sqrt{\alpha_x^2 - 4 \Gamma t_x} \right) .
\end{align}

Similarly, we find that the kinetic term is diagonal in the basis of the two states, with
\begin{align}
  \Braket{\Psi'^1_{0,\pm} | \alpha_y k_y \tau_y | \Psi'^1_{0,\pm}}
  &= \pm v_{\trm{F}} k_y ,
    \notag \\
  \Braket{\Psi'^1_{0,+} | \alpha_y k_y \tau_y | \Psi'^1_{0,-}}
  &= 0 ,
\end{align}
where $v_{\trm{F}} = \alpha_y$. However, unlike on the edge $s=0$,
here we find that  all expectation values of the Zeeman term are
exactly zero
\begin{align}
  \Braket{\Psi'^1_{0,\pm} |
  \Delta_{\trm{z}} \sigma_x
  | \Psi'^1_{0,\pm}} &=
    0 ,
    \notag \\
  \Braket{\Psi'^1_{0,+} |
  \Delta_{\trm{z}} \sigma_x
  | \Psi'^1_{0,-}} &=
    0 .
\end{align}
We note that this feature is independent of the orientation of the
Zeeman field and is crucial to generate four corner states. Taking
into account the proximity induced superconductivity effect, we can
now express the effective low-energy Hamiltonian as
\begin{align}
  \mc{H}'_{\trm{eff}, 1}(k_{y})
  = v_{\trm{F}} k_{y} \rho_z + \Delta_{\trm{sc}} \eta_y \rho_y ,
\end{align}
which we immediately identify with the topologically trivial regime.

\begin{figure*}[t]
  \centering
  \includegraphics[width=1.50\columnwidth]{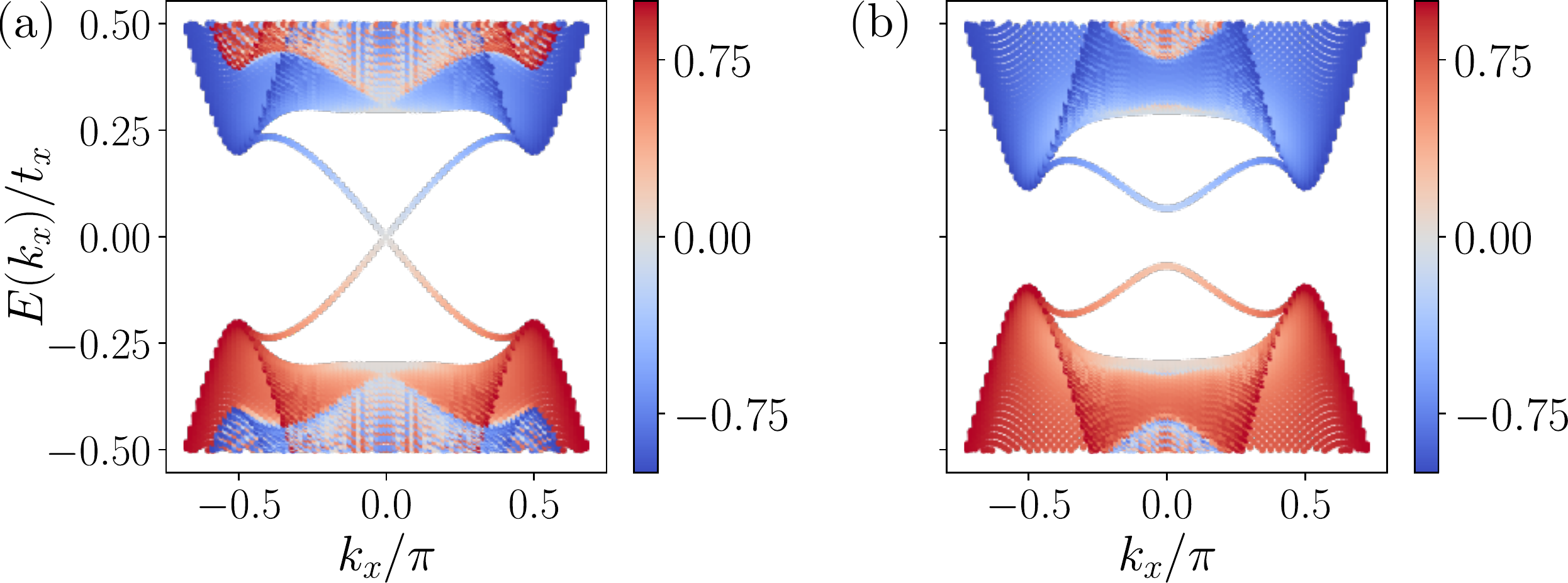}
  \caption{Numerical calculation of the spin polarization of the
    eigenstates of $\mc{H}'$ in a geometry with OBC (PBC) along  the $x$
    ($y$) axis. The color corresponds to
    $\Braket{\Psi_{j} (k_y) | S_{\parallel} | \Psi_{j} (k_y) }$, where
    $\Ket{\Psi_{j} (k_y)}$ is the $j$-th eigenstate of $\mc{H'}$. (a)
    When $\Delta_{\trm{sc}} = 0$, the edge spectrum remains gapless
    and the spin polarization of the states is zero at $k_y = 0$. (b)
    When $\Delta_{\trm{sc}} = 0.1 t_x$, the edge spectrum is gapped
    out and the spin polarization becomes non-zero even at $k_y =
    0$. Nevertheless, the sign of the spin polarization remains the
    same as in (a). All the remaining parameters are the same as in
    Fig.~\ref{fig:soti_mirror}(b).}
  \label{fig:polar_2}
\end{figure*}

In order to find the spin polarization of the edge states along the edge $s = 1$, we solve the problem by explicitly taking into account the kinetic and Zeeman terms. The new problem is described by the Hamiltonian $\mc{H}'_{0}(k_x, k_y) + \Delta_{\trm{z}} \sigma_x$.  To  lowest order in $k_y$ and $\Delta_{\trm{z}}$, the corresponding solutions can be expressed as 
\begin{widetext}
\begin{align}
  \label{eq:psi0_inversion_3}
  \tilde{\Psi}'^1_{0,+}(x)
  = &
    \sum\limits_{\lambda=\pm}
    \lambda
    \frac{e^{-x / \xi_{1,\lambda}}}{\mc{N}}
    \left[ 1 + f_{1,\lambda}, 1 + f_{1,\lambda},
    i (1 - f_{1,\lambda}), i (f_{1,\lambda} - 1) \right]^{T}
    \notag \\
  + &
    \sum\limits_{\lambda=\pm}
    \lambda
    \frac{e^{-x / \xi_{2,\lambda}}}{\mc{N}}
    \left[ 1 + f_{2,\lambda},-(1 + f_{2,\lambda}),
    i (1 - f_{2,\lambda}), i (1 - f_{2,\lambda}) \right]^{T},
    \notag \\
  \tilde{\Psi}'^1_{0,-}(x)
  = &
    \sum\limits_{\lambda=\pm}
    \lambda
    \frac{e^{-x / \xi_{1,\lambda}}}{\mc{N}}
    \left[ 1 + f_{1,\lambda}, 1 + f_{1,\lambda},
    i (1 - f_{1,\lambda}), i (f_{1,\lambda} - 1) \right]^{T}
    \notag \\
  - &
    \sum\limits_{\lambda=\pm}
    \lambda
    \frac{e^{-x / \xi_{2,\lambda}}}{\mc{N}}
    \left[ 1 + f_{2,\lambda},-(1 + f_{2,\lambda}),
    i (1 - f_{2,\lambda}), i (1 - f_{2,\lambda}) \right]^{T},
    \notag \\
\end{align}
\end{widetext}
with $f_{j,\pm} = (-1)^{j} \xi_{j,\pm}\alpha_y k_y / (2 \alpha_x)$,
$\tilde{\xi}^{-1}_{1,\pm} = \left( \alpha_x \pm \sqrt{\alpha_x^2 - 4
    (\Gamma + \Delta_{\trm{z}}) t_x} \right) / (2 t_x)$, and
$ \tilde{\xi}^{-1}_{2,\pm} = \left( \alpha_x \pm \sqrt{\alpha_x^2 - 4
    (\Gamma - \Delta_{\trm{z}}) t_x} \right) / (2 t_x)$. We verify
that the new solutions reproduce correctly
Eq.~\eqref{eq:psi0_inversion_2} in the limit $\Delta_{\trm{z}} = 0$
and $k_y = 0$. We also find that under the effect of the Zeeman field,
the condition of the existence of the edge modes is modified to
$\Gamma > |\Delta_\trm{z}|$, such that the topological phase becomes
smaller when the Zeeman field increases. Moreover, the polarization of
the edge states is defined by the following expression:
\begin{widetext}
\begin{align}
    \Braket{\tilde{\Psi}'^1_{0,\pm} |
    S_{\parallel}
    | \tilde{\Psi}'^1_{0,\pm}} = \pm 4
    \frac
    {\sum_{j} \mathfrak{Re} \left[
    f_{j, +} \xi_{j,+} +
    f_{j, -} \xi_{j,-} -
    4 f_{j, +} f_{j, -} / (\xi^{-1}_{j,+} + \xi^{-1*}_{j,-}) \right]}
    {\sum_{j} \mathfrak{Re} \left[ \xi_{j,+} + \xi_{j,-} -
    4 / (\xi^{-1}_{j,+} + \xi^{-1*}_{j,-}) \right]}.
\end{align}
\end{widetext}
As expected, the parallel component of the spin polarization is zero
for $\Delta_{\trm{z}} = 0$ independently of the value of $k_y$. The
same is true for $k_y = 0$, independently of the value of
$\Delta_{\trm{z}}$. These features are confirmed numerically in
Fig.~\ref{fig:polar_2}(a), where we calculate the spin polarization of
the eigenstates of $\mc{H}'$ in a geometry with the OBC (PBC) along the $x$
($y$) axis as a function of the momentum $k_y$. Moreover, we find that
  the spin polarization changes smoothly as a function of
  $\Delta_{\trm{z}}$. We also note that both perpendicular
  components of the spin polarization are exactly zero. 

If the superconducting term is taken into account, the spectrum of states $\Ket{\Psi'^{1}_{0,\pm}}$ acquires a finite gap. Nevertheless,
  we expect that the spin polarization of the edge states keeps the
  same sign for different values of the ratio
  $\Delta_{\trm{z}} / \Delta_{\trm{sc}}$ across the entire phase
  transition. This is confirmed by numerical calculations presented in
  Fig.~\ref{fig:polar_2}(b). We also note that for
  $\Delta_{\trm{sc}} \neq 0$, the edge states become spin polarized
  even at $k_y = 0$.

\begin{figure}[t]
  \centering
  \includegraphics[width=.99\columnwidth]{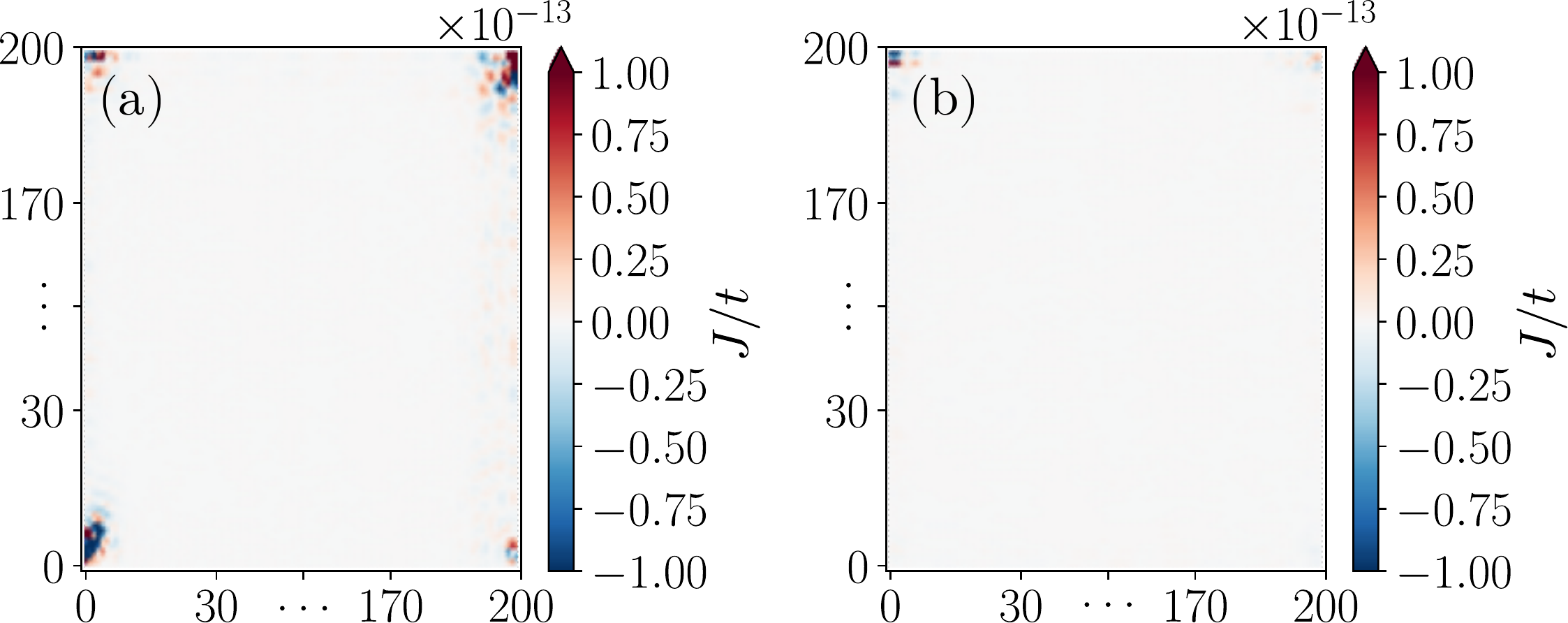}
  \caption{Numerical calculation of the current $J$ through the STM
    tip that is polarized perpendicularly to the Zeeman field along
    the $y$ axis (a) in the topological phase with
    $\Delta_{\trm{z}} = 1.75 \Delta_{\trm{sc}}$ and (b) in the trivial
    phase with $\Delta_{\trm{z}} = 0.25 \Delta_{\trm{sc}}$. The
    remaining parameters of the simulations are the same as in
    Fig.~\ref{fig:soti_mirror_stm}. The perpendicular component of the
    spin polarization is trivially zero everywhere except at the four
    corners of the system.}
  \label{fig:soti_mirror_stm_perp}
\end{figure}

\subsubsection{Quadrupolar structure of the spin polarization}

Finally, by considering the effect of the inversion symmetry that maps
$\vec{k}$ to $-\vec{k}$, we relate the description of the edge $s=2$
($s=3$) to the one of the edge $s=0$ ($s=1$). More specifically, we
use that $I \mc{H}'(\vec{k}) I^{-1} = \mc{H}'(-\vec{k})$, where
$I = \tau_z$, which implies that the mass terms at opposite edges have
opposite signs: $m_0 = -m_2 = \Delta_{\trm{z}}$ and $m_1 = -m_3 =
0$. This allows us to see that the total boundary of the system is
composed of two pairs of effective Rashba wires forming an alternating
pattern, only one of which is affected by the Zeeman field. As a
result, in the topological regime
$\Delta_{\trm{z}} > \Delta_{\trm{sc}}$, the system is described by the
parallel component of the spin polarization flipping its sign from one
edge to another, resulting in a quadrupolar spin structure, which can
indeed be observed in Fig.~\ref{fig:soti_mirror_stm} of the main
text. Such a feature is directly associated with the emergence of MCSs
and allows one to probe the topological phase transition which occurs
in the system. We also verify numerically (see
Fig.~\ref{fig:soti_mirror_stm_perp}) that the perpendicular components
of the spin polarization are trivially zero everywhere except at the
four corners of the system, where they acquire some finite value as a
result of the broken translation symmetry along
the edge.\\

\section{Quadrupolar moment}

In the main text, we demonstrated that the structure of the edge spin polarization allows one to detect the topological phase transition in SOTSCs with broken time-reversal symmetry. In particular, we found that in SOTSCs 
hosting a pair of MCSs at two opposite corners the sign of the
spin polarization perpendicular to the Zeeman field of low-energy states changes on every edge. This feature has been denoted as quadrupolar structure of the spin polarization. Similarly, we observed that a SOTSC which hosts a MCS at each of the four corners is described by a quadrupolar structure of the
spin polarization parallel to the applied Zeeman field.

\begin{figure}[t]
  \centering
  \includegraphics[width=.99\columnwidth]{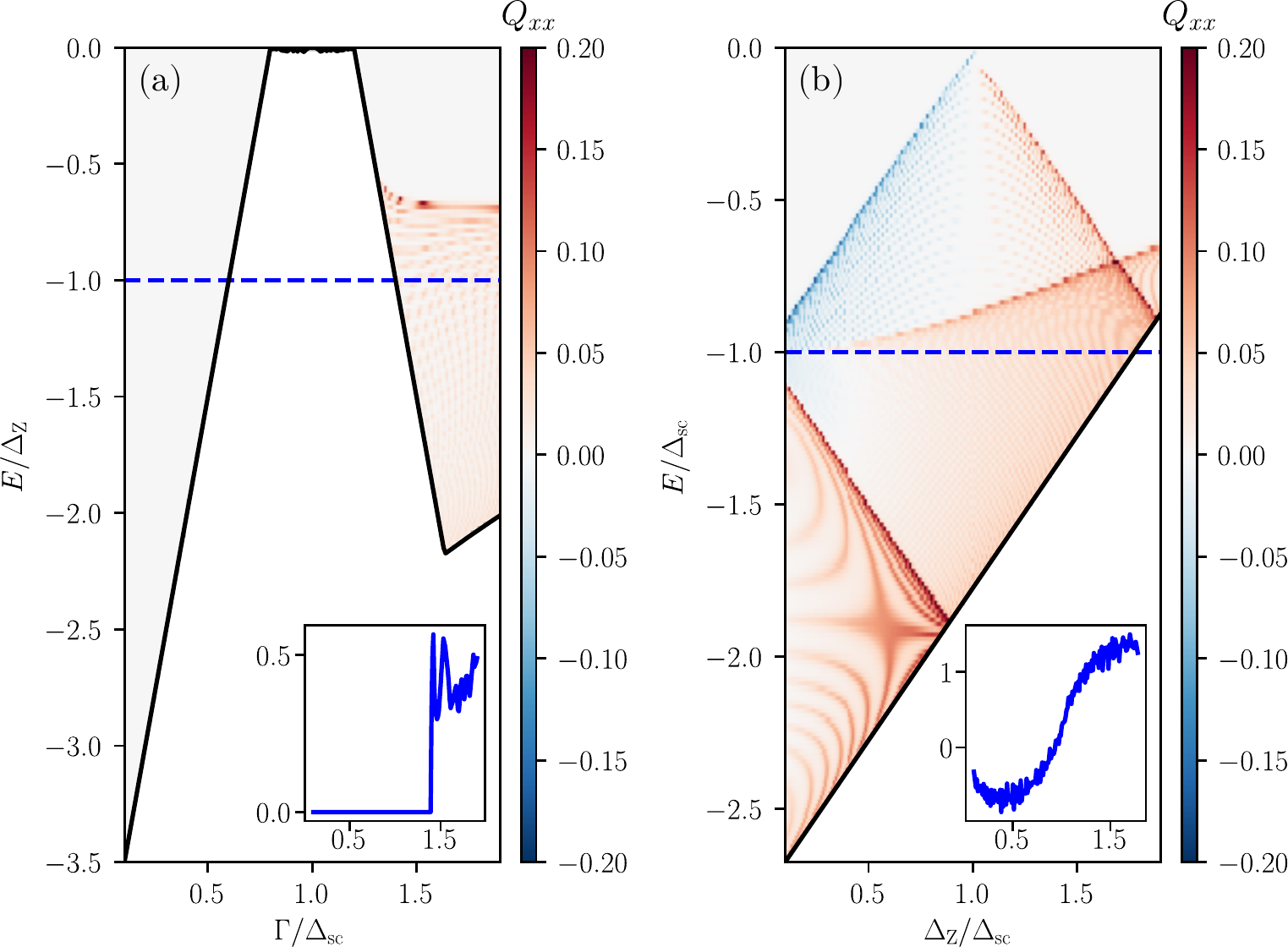}
  \caption{Energy-resolved value of the quadrupolar moment $Q_{xx}$ evaluated across the topological phase transition for the two models considered in the main text. 
  Insets show the contribution summed from the blue lines up to the chemical potential. Black lines separate the contribution coming from the edges (shown) and the bulk (not shown). (a) The model is described by the Hamiltonian $\mc{H}$ and $Q_{xx}$ is calculated as a
    function of $\Gamma / \Delta_{\trm{sc}}$ with $\Delta E = 0.02 \Delta_{\trm{z}}$. All the remaining parameters are the same as in Fig.~\ref{fig:soti_inversion_stm}. The trivial (topological) phase corresponds to zero
    (finite) value of the quadrupolar moment. (b) The model is
    described by the Hamiltonian $\mc{H}'$ and $Q_{xx}$ is calculated
    as a function of $\Delta_{\trm{z}} / \Delta_{\trm{sc}}$ with $\Delta E = 0.015 \Delta_{\trm{sc}}$. All the
    remaining parameters are the same as in
    Fig.~\ref{fig:soti_mirror_stm}. The spin polarization of the $x$-edge states changes sign at the phase transition, while for $y$-edge states it slowly increases with $\Delta_{\trm{z}}$. As a consequence, the integrated value of $Q_{xx}$ crosses zero close to the topological phase transition point.}
  \label{fig:Qxx_integrated}
\end{figure}

Here, we further analyze the quadrupolar structure of the spin polarization by introducing the energy-resolved quadrupolar tensor $Q_{\mu \nu}$ associated with the STM current $J$. Assuming that the sample is a perfect square and by placing the origin of coordinates in the square center we can define the following quantity:
\begin{align}
  Q_{\mu \nu}(E) = \sum\limits_{i}
  \left[ 2 r_{\mu}(i) r_{\nu}(i) - \mb{I}_{\mu \nu} \right] J(\vec{r}_i, E) / N .
\end{align}
Here $r_{\mu}(i)$ denotes the position of the site with the lattice index $i$ and $\mb{I}$ is the $2 \times 2$ identity matrix. The sum over $i$ runs over all the sites of the system and $N$ denotes the total number of sites. The current $J(\vec{r}_i, E)$ refers to the contribution defined in Eq.~\eqref{eq:current_sup} for a small window $[E-\Delta E/2, E+\Delta {E}/2]$.
We calculate this quantity in both models across the phase
transition. We also focus only on the diagonal component $Q_{xx} = -Q_{yy}$ of the quadrupolar tensor with the off-diagonal components being trivially zero. 

First, we consider the model presented in Section ``SOTSCs with two corner states" [see Fig.~\ref{fig:Qxx_integrated}(a)]. We assume that the STM tip polarization is perpendicular to the Zeeman field and calculate $Q_{xx}$ as a function of $\Gamma / \Delta_{\trm{sc}}$. We find that in the topologically trivial phase, the quadrupolar moment is exactly zero, since no available edge states are present in the considered energy window. In the topological phase, on the contrary, $Q_{xx}$ is positive that can be associated with the quadrupolar structure of spin polarization at the edges.

Similarly, we consider the model presented in Section ``SOTSCs with four corner states"
with the STM tip polarization being parallel to the Zeeman field [see Fig.~\ref{fig:Qxx_integrated}(b)]. We clearly distinguish the contribution coming from different edges: the energy of the $x$-edge states increases with $\Delta_{\trm{z}}$ until it reaches zero at the critical point $\Delta_{\trm{z}} = \Delta_{\trm{sc}}$, after which it starts decreasing again;  a large quadrupolar moment flips sign at the phase transition. For $\Delta_{\trm{z}} > \Delta_{\trm{sc}}$, we also observe the emergence of edge states at lower energies, which live on the $x$-edge and are described by the spin polarization of an opposite sign. At the same time, the energy of the $y$-edge states as well as their quadrupolar moment slowly increases with $\Delta_{\trm{z}}$ without flipping its sign. As a result, when integrated over the energy $E$, the quadrupolar moment changes sign at the topological phase transition.

We notice that the precise value of $Q_{xx}$ depends strongly on the energy $E$.  Nevertheless, the quadrupolar structure, namely the sign change of the spin polarization on the neighboring edges of the system,
is typical for topological phases close to the phase boundary. Hence, the quadrupolar structure of the spin polarization remains a prominent probe of the SOTSCs topology as long as the effective low-energy description stays valid. 

Finally, we note that the quadrupolar tensor $Q_{\mu \nu}(E)$ is less suitable for direct experimental observation as it requires integration of the current signal across the entire sample. Moreover, the values of $Q_{\mu \nu}(E)$ depend on the sample geometry such as size and shape, as well as on the choice of coordinate origin in the definition of $Q_{\mu \nu}$, since, in general, the total spin polarization and dipole moments are non-zero.

\begin{figure}[t]
  \centering
  \includegraphics[width=.99\columnwidth]{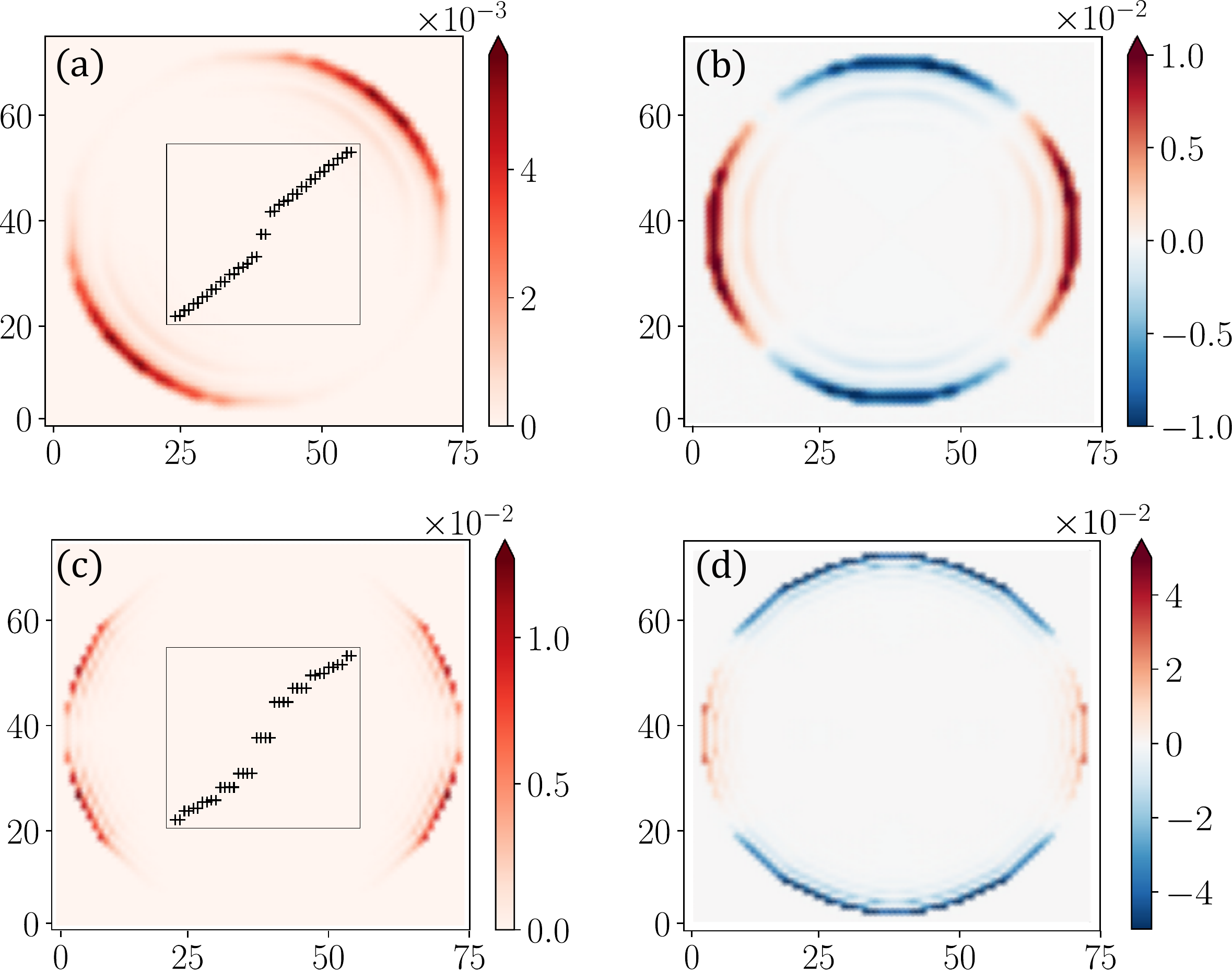}
  \caption{(a) Numerical calculation of the MCS wavefunctions and (b) the total spin polarization of 40 states below the Fermi level in the topological phase of the system described by the Hamiltonian $\mc{H}$ in a disk geometry. The parameters used are $\Gamma = 2 \Delta_{\trm{sc}} = \alpha = 0.5 t$, $\mu_0 = 0$, $\Delta_{\trm{z}} = 0.05t$, and $\theta_{\trm{z}} = \pi / 4$. (c,d) The same calculation as before but the system is now described by the Hamiltonian $\mc{H}'$ with $\Gamma = t_y = t_x$, $\alpha_x = \alpha_y = 0.3 t_x$, $\mu_0 = 0$, and $\Delta_{\trm{z}} = 2 \Delta_{\trm{sc}} = 0.1 t_x$. As expected, in both cases, we observe MCSs,  emphasizing that the shape of the sample does not play a substantial role. The corresponding spin polarization shows the quadrupole structure.}
  \label{fig:circle}
\end{figure}

\begin{figure*}[!t]
  \centering
  \includegraphics[width=1.99\columnwidth]{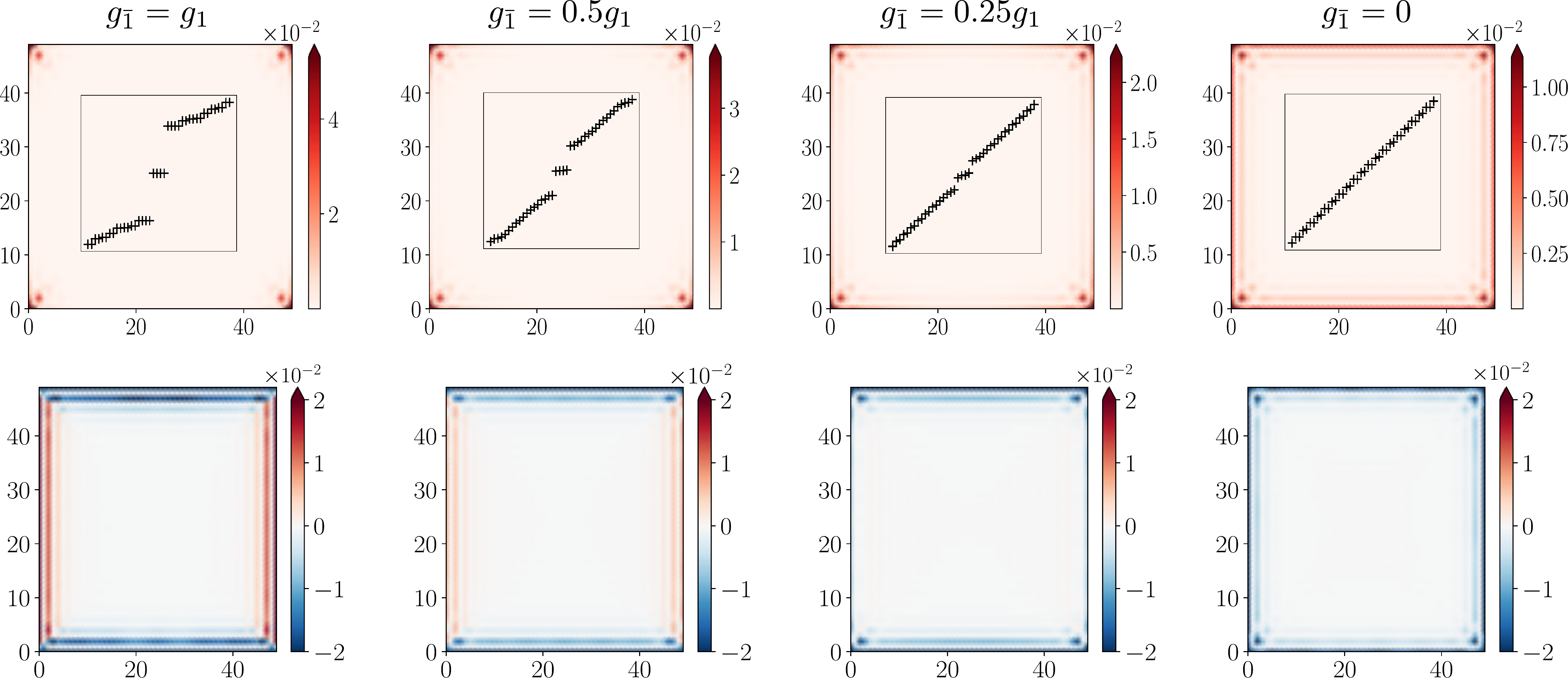}
  \caption{(first row) Numerical calculation of the MCS wavefunctions and (second row) the total spin polarization of 40 states below the Fermi level in the topological phase of a system described by the Hamiltonian $\mc{H}'$ for different values of the $g$-factor ratio $g_{\bar{1}} / g_{1}$ (different columns). The remaining parameters are the same as in Fig.~\ref{fig:soti_mirror}(b) of the main text. We confirm that the inversion symmetry between two bands, which gets broken if $g$-factors are different, is indeed not needed. The MCSs survive even if the $g$-factors are substantially different.}
  \label{fig:mirror_gfactor}
\end{figure*}

\section{Stability of  quadrupolar polarization}

We also provide additional numerical
results, demonstrating the stability of the quadrupolar polarization
feature for the two SOTSC classes considered in the main text. To do
this, we simulate the system in a disc geometry, which does not have
any well defined edges and corners. Nevertheless, MCSs still emerge in
such a geometry. Their position is unambiguously determined by the
symmetries of the system, namely, the inversion symmetry $I$ of the
Hamiltonian $\mc{H}$ from the section ``SOTSCs with two corner states"
and the in-plane anisotropy of the Hamiltonian $\mc{H}'$ from the
section ``SOTSCs with four corner states" of the main text. In
Figs.~\ref{fig:circle}(a) and~(b), we show that, in the model
described by $\mc{H}$, two corner states emerge at two opposite
extremities of the disc, aligned with the direction of the Zeeman
field. The perpendicular component of the spin polarization changes
sign at four equal-sized quadrants delimited by the Zeeman field
vector and the vector normal to it. Similarly, in
Figs.~\ref{fig:circle}(c) and~(d), we show that four corner states
emerge in the model described by $\mc{H}'$, dividing the disk into
four quadrants. The parallel component of the spin polarization has
opposite signs in neighboring quadrants, while the perpendicular
component remains trivially zero everywhere. The size of the quadrants
is determined by the ratio $\Delta_{\trm{z}} / \Delta_{\trm{sc}}$,
such that in the limit $\Delta_{\trm{z}} \ll \Delta_{\trm{sc}}$ the
corner states merge pairwise at the top and bottom extremities of the
disk.

Additionally, having in mind an experimental realization in quantum wells~\cite{koenig2007, molenkamp2009, wangzhang2008,gfactor1,gfactor2,gfactor3,gfactor4} of our model with four corner states described by the Hamiltonian $\mc{H}'$, we study how the topological phase diagram and the result of the STM measurement vary as a function of $g$-factors of the electron and hole bands, denoted by $g_{1}$ and $g_{\bar{1}}$, respectively. The result of such a calculation is presented in Fig.~\ref{fig:mirror_gfactor}. We find that the topological phase and, as a result, the quadrupolar polarization feature are stable even for a strong $g$-factor anisotropy. However, when $g_{\bar{1}}$ becomes of the order of $0.25 g_{1}$ (and vice versa), a phase transition occurs leading to the closing of the bulk gap, accompanied by the disappearance of the corner states as well as of the quadrupolar structure of the spin polarization.



\begin{thebibliography}{1}


\bibitem{TI_KaneMele2005_1}
  C. L. Kane and E. J. Mele,
  \prl\ {\bf 95}, 226801 (2005).
	
\bibitem{TI_KaneMele2005_2}
  C. L. Kane and E. J. Mele,
  \prl\ {\bf 95}, 146802 (2005).
	
\bibitem{TI_BernevigHughesZhang2006}
  B. A. Bernevig, T. L. Hughes, and S. Zhang,
  Science {\bf 314}, 5806 (2006).

\bibitem{TIRev_HasanKane2010}
  M. Z. Hasan and C. L. Kane,
  Rev. Mod. Phys. {\bf 82}, 3045 (2010).
	
\bibitem{TIRev_QiZhang2011}
  X.-L. Qi and S.-C. Zhang,
  Rev. Mod. Phys. {\bf 83}, 1057 (2011).

\bibitem{TIRev_SatoAndo2017}
  M. Sato and Y. Ando,
  Rep. Prog. Phys. \tbf{80}, 076501 (2017).

\bibitem{TIRev_WangZhang2017}
  J. Wang and S.-C. Zhang,
  Nat. Mat. \tbf{16}, 1062-1067 (2017).

\bibitem{TIRev_Wen2017}
  X.-G. Wen,
  Rev. Mod. Phys. \tbf{89}, 41004 (2017).

\bibitem{QComp_FreedmanKitaevLarsenWang2003}
  M. H. Freedman, A. Kitaev, M. J. Larsen, and Z. Wang,
  Bull. Amer. Math. Soc. \tbf{40}, 31 (2003).

\bibitem{QComp_Kitaev2003}
  A. Kitaev,
  Annals Phys. \tbf{303}, 2 (2003).

\bibitem{QComp_SternLindner2013}
  A. Stern and N. H. Lindner,
  Science \tbf{339}, 6124 (2013).
  
\bibitem{HOTI_BenalcazarBernevigHughes2017}
  W. A. Benalcazar, B. A. Bernevig, and T. L. Hughes,
  Science \tbf{357}, 6346 (2017).

\bibitem{HOTI_BenalcazarBernevigHughes2017_2}
  W. A. Benalcazar, B. A. Bernevig, and T. L. Hughes,
  \prb\ \tbf{96}, 245115 (2017).

\bibitem{HOTI_SongFangFang2017}
  Z. Song, Z. Fang, and C. Fang,
  \prl\ \tbf{119}, 246402 (2017).

\bibitem{HOTI_LangbehnPengTrifunovicEtAl2017} J. Langbehn, Y.
  Peng, L. Trifunovic, F. von Oppen, and P. W. Brouwer,
  Phys. Rev. Lett. \tbf{119}, 246401 (2017).
    
\bibitem{PetersonBenalcazarHughesBahl2018} C. W. Peterson,
  W. A. Benalcazar, T. L. Hughes, and G. Bahl,
  Nature \tbf{555}, 346 (2018).

\bibitem{ImhofEtAl2018} S. Imhof, C.  Berger,
  F. Bayer, J. Brehm, L. W. Molenkamp, T.  Kiessling, F. Schindler,
  C. H. Lee, M. Greiter, T.  Neupert, and R. Thomale,
  Nat. Phys. \tbf{14}, 925 (2018).

\bibitem{MittalEtAl2019} S. Mittal, V. V. Orre, G. Zhu, M. A. Gorlach,
  A. Poddubny, and M. Hafezi,
  Nat. Photon. \tbf{10}, 1038 (2019).

\bibitem{ChenEtAl2019}
  X.-D. Chen, W.-M. Deng, F.-L. Shi, F.-L. Zhao, M. Chen, and
  J.-W. Dong, Phys. Rev. Lett. \tbf{122}, 233902 (2019).
  
\bibitem{HassanKunstEtAl2019} A. E. Hassan, F. K. Kunst, A. Moritz,
  G. Andler, E. J. Bergholtz, and M. Bourennane,
  Nat. Photon. \tbf{13}, 697 (2019).

\bibitem{SerraGarciaEtAl2018} M. Serra-Garcia, V. Peri, R. S\"{u}sstrunk, O.  R. Bilal,
  T. Larsen, L. G. Villanueva, and S. D. Huber, Nature \tbf{555}, 342
  (2018).

\bibitem{XueYangEtAl2019} H. Xue, Y. Yang, F. Gao, Y. Chong, and
  B. Zhang, Nature Materials \tbf{18}, 108 (2019).

\bibitem{SchindlerEtAl2018_1} F. Schindler, A. M. Cook, M.
  G. Vergniory, Z. Wang, S. S. P. Parkin, B. A. Bernevig,
  and T. Neupert, Science Advances \tbf{4}, 6 (2018).
  
\bibitem{SchindlerEtAl2018_2} F. Schindler, Z. Wang, M.
  G. Vergniory, A. M. Cook, A. Murani, S. Sengupta, A.
  Y. Kasumov, R. Deblock, S. Jeon, I. Drozdov, H.
  Bouchiat, S. Gu\"{e}ron, A. Yazdani, B. A. Bernevig, and T.
  Neupert, Nature Physics \tbf{14}, 918 (2018).

\bibitem{WeinerNiLiAluKhanikaev2019} X. Ni, M. Weiner, A. Al\'{u},
  and A. B. Khanikaev, Nature Materials \tbf{18}, 113 (2019).

\bibitem{ZhangXieEtAl2019}
  X. Zhang, B.-Y. Xie, H.-F. Wang, X. Xu, Y. Tian, J.-H. Jiang,
  M.-H. Lu, Y.-F. Chen,
  Nat. Comm. \tbf{10}, 5331 (2019).
  
\bibitem{ShiozakiSato_2014}
  K. Shiozaki and M. Sato,
  Phys. Rev. B \tbf{90}, 165114 (2014).
  
\bibitem{GeierEtAl_2018}
  M. Geier, L. Trifunovic, M. Hoskam, and P. W. Brouwer,
  Phys. Rev. B \tbf{97}, 205135 (2018).
  
\bibitem{Khalaf_2018}
  E. Khalaf,
  Phys. Rev. B \tbf{97}, 205136 (2018).
  
\bibitem{Trifunovic_2020}
  L. Trifunovic and P. W. Brouwer,
  arXiv:2003.01144.
  
\bibitem{Zhu_2018}
  X. Zhu,
  Phys. Rev. B \tbf{97}, 205134 (2018).
  
\bibitem{VopezEtAl_2019}
  Y. Volpez, D. Loss, and J. Klinovaja,
  Phys. Rev. Lett. \tbf{122}, 126402 (2019).
  
\bibitem{LaubscherEtAl_2019}
  K. Laubscher, D. Loss, and J. Klinovaja,
  Phys. Rev. Research \tbf{1}, 032017 (2019).
  
\bibitem{AhnYang_2020}
  J. Ahn and B.-J. Yang,
  Phys. Rev. Research \tbf{2}, 012060(R) (2020).

\bibitem{WangLinHughes_2018}
  Y. Wang, M. Lin, and T. L. Hughes,
  Phys. Rev. B \tbf{98}, 165144 (2018).

\bibitem{LiuHeNori_2018}
  T. Liu, J. J. He, and F. Nori,
  Phys. Rev. B \tbf{98}, 245413 (2018).
  
\bibitem{FrancaEfremovFulga_2019}
  S. Franca, D. V. Efremov, and I. C. Fulga,
  Phys. Rev. B \tbf{100}, 075415 (2019).
  
\bibitem{Yan_2019}
  Z. Yan,
  Phys. Rev. B \tbf{100}, 205406 (2019).
  
\bibitem{ZhangColeWeDasSarma_2019}
  R.-X. Zhang, W. S. Cole, X. Wu, and S. Das Sarma,
  Phys. Rev. Lett. \tbf{123}, 167001 (2019).
  
\bibitem{Zhu_2019}
  X. Zhu,
  Phys. Rev. Lett. \tbf{122}, 236401 (2019).
  
\bibitem{LaubscherEtAl_2020}
  K. Laubscher, D. Loss, and J. Klinovaja,
  Phys. Rev. Research \tbf{2}, 013330 (2020).

\bibitem{WuLiuThomaleLiu_2019}
  X. Wu, X. Liu, R. Thomale, C.-X. Liu,
  arXiv:1905.10648.

\bibitem{WuEtAl_2019}
  Y.-J. Wu, J. Hou, Y.-M. Li, X.-W. Luo, and C. Zhang,
  arXiv:1905.08896.
  
\bibitem{ZhangCalzonaTrauzettel_2020}
  S.-B. Zhang, A. Calzona, and B. Trauzettel,
  arXiv:2003.04053.
  
\bibitem{kramers_YanSongWang_2018}
  Z. Yan, F. Song, and Z. Wang,
  Phys. Rev. Lett. \tbf{121}, 096803 (2018).
  
\bibitem{kramers_WangLiuLuZhang_2018}
  Q. Wang, C.-C. Liu, Y.-M. Lu, and F. Zhang,
  Phys. Rev. Lett. \tbf{121}, 186801 (2018).
  
\bibitem{kramers_HsuEtAl_2018}
  C.-H. Hsu, P. Stano, J. Klinovaja, and D. Loss,
  Phys. Rev. Lett. \tbf{121}, 196801 (2018).
  
\bibitem{kramers_ChenLiuXuLiu_2019}
  L. Chen, B. Liu, G. Xu, X. Liu,
  	arXiv:1909.10402.
  
\bibitem{kramers_HsuColeZhangSau_2019}
  Y.-T. Hsu, W. S. Cole, R.-X. Zhang, and J. D. Sau,
  arXiv:1904.06361.  
  
\bibitem{ZitkoLimLopezAquado2015}
  R. Zitko, J. S. Lim, R. L\'{o}pez, and R. Aguado,
  Phys. Rev. B \tbf{91}, 045441 (2015).

\bibitem{PradaEtAl2020}
  E. Prada, P. San-Jose, M. W. A. de Moor, A. Geresdi, E. J. H. Lee, J. Klinovaja,
  D. Loss, J. Nygard, R. Aguado, and L. P. Kouwenhoven,
  arXiv:1911.04512.
  
\bibitem{abs} C. Reeg, O. Dmytruk, D. Chevallier, D. Loss, and J. Klinovaja,
Phys. Rev. B {\bf 98}, 245407 (2018).

\bibitem{abs1} A. Ptok, A. Kobiaka, and T. Domanski, Phys. Rev. B {\bf 96}, 195430 (2017).

\bibitem{abs2} F. Setiawan, C.-X. Liu, J. D. Sau, and S. Das Sarma, Phys. Rev. B {\bf 96}, 184520 (2017).

\bibitem{abs3}  C. Moore, T. D. Stanescu, and S. Tewari, Phys. Rev. B 97, 165302 (2018).

\bibitem{abs4} A. Vuik, B. Nijholt, A. R. Akhmerov, and M. Wimmer, SciPost Phys. \tbf{7}, 061 (2019).

\bibitem{abs5} J. Avila, F. Penaranda, E. Prada, P. San-Jose, and R. Aguado, Communications Physics \tbf{2}, 133 (2019).

\bibitem{abs6} E. B. Hansen, J. Danon, and K. Flensberg, Phys. Rev. B {\bf 97,} 041411 (2018).
\bibitem{abs7} F. Penaranda, R. Aguado, P. San-Jose, and E. Prada, Phys. Rev. B \tbf{98}, 235406 (2018).
\bibitem{abs8} G. Kells, D. Meidan, and P. W. Brouwer, Phys. Rev. B {\bf 86}, 100503 (2012).
\bibitem{abs9} C. Fleckenstein, F. Dominguez, N. Traverso Ziani, and B. Trauzettel, Phys. Rev. B {\bf 97}, 155425 (2018).

\bibitem{sch} C. J\"{u}nger, R. Delagrange, D. Chevallier, S. Lehmann, K.A. Dick, C. Thelander, J. Klinovaja, D. Loss, A. Baumgartner, and C. Sch\"{o}nenberger, 
Phys. Rev. Lett. {\bf 125}, 017701 (2020).
  
\bibitem{AlspaughSheehyGoerbigSimon2020}
  D. J. Alspaugh, D. E. Sheehy, M. O. Goerbig, and P. Simon,
  Phys. Rev. Research \tbf{2}, 023146 (2020).
  
\bibitem{ZhangTrauzettel_2020}
  S.-B. Zhang and B. Trauzettel,
  Phys. Rev. Research \tbf{2}, 012018 (2020).
  
\bibitem{SzumniakEtAl2017}
  P. Szumniak, D. Chevallier, D. Loss, and J. Klinovaja,
  Phys. Rev. B \tbf{96}, 041401(R) (2017).
  
\bibitem{SerinaLossKlinovaja2018}
  M. Serina, D. Loss, and J. Klinovaja,
  Phys. Rev. B \tbf{98}, 035419 (2018).

\bibitem{ThakurathiEtAl2020}
  M. Thakurathi, D. Chevallier, D. Loss, J. Klinovaja,
  Phys. Rev. Research \tbf{2}, 023197 (2020).
  
\bibitem{MullerEtAl2020}
  N. M\"{u}ller, D. M. Kennes, J. Klinovaja, D. Loss, and H. Schoeller,
  Phys. Rev. B \tbf{101}, 155417 (2020).

\bibitem{JozwiakEtAl2016}
  C. Jozwiak, J. A. Sobota, K. Gotlieb, A. F. Kemper,
  C. R. Rotundu, R. J. Birgeneau, Z. Hussain, D.-H. Lee,
  Z.-X. Shen, and A. Lanzara,
  Nature Com., \tbf{7}, 13143 (2016).

\bibitem{JeonEtAl2017}
  S. Jeon, Y. Xie, J. Li, Z. Wang, B. A. Bernevig, and A. Yazdani,
  Science \tbf{358}, 6364 (2017).

\bibitem{JackEtAl2019}
  B. J\"{a}ck, Y. Xie, J. Li, S. Jeon, B. A. Bernevig, and A. Yazdani,
  Science \tbf{364}, 6447 (2019). 
  
\bibitem{KimEtAl2019}
  H. Kim, A. Palacio-Morales, T. Posske, L. R\'{o}zsa, K. Palot\'{a}s, L. Szunyogh, M.
  Thorwart, and R. Wiesendanger,
  Science Advances \tbf{11}, 5251 (2018). 

\bibitem{RubyAl2019}
  M. Ruby, B. W. Heinrich, Y. Peng, F. von Oppen, and K. J. Franke,
  Nano Lett. \tbf{17}, 4473 (2017). 
  
\bibitem{Jackiw1Rebbi976}
  R. Jackiw and C. Rebbi,
  Phys. Rev. D \tbf{13}, 3398 (1976).
	
\bibitem{JackiwSchrieffer1981}
  R. Jackiw and J. Schrieffer,
  Nucl. Phys. B \tbf{190}, 253 (1981).  
  
\bibitem{SM} See Supplemental Material at \dots for (1) microscopic
  details of the models; (2) details on the numerical methods; (3)
  details on the analytical calculations of the edge states and the
  spin polarization; (4) a study of the quadrupolar moment across the
  phase transition and (5) the stability of the quadrupolar features
  against external perturbations.
  
\bibitem{NilssonEtAl2008}
  J. Nilsson, A. R. Akhmerov, and C. W. J. Beenakker,
  Phys. Rev. Lett. \tbf{101}, 120403 (2008).
  
\bibitem{FuKane2009}
  L. Fu and C. L. Kane,
  Phys. Rev. B \tbf{79}, 161408(R) (2009).
  
\bibitem{LutchynEtAl2010}
  R. M. Lutchyn, J. D. Sau, and S. Das Sarma,
  Phys. Rev. Lett. \tbf{105}, 077001 (2010).
  
\bibitem{OregEtAl2010}
  Y. Oreg, G. Refael, and F. von Oppen,
  Phys. Rev. Lett. \tbf{105}, 177002 (2010).
  
\bibitem{AliceaEtAl2010}
  J. Alicea
  Phys. Rev. B \tbf{81}, 125318 (2010).

\bibitem{koenig2007}
M.~K\"{o}nig, S.~Wiedmann, C.~Br\"{u}ne, A.~Roth, H.~Buhmann, L. W.~Molenkamp, X.~Qi, S.~Zhang, Science \textbf{318}, 5851 (2007).

\bibitem{molenkamp2009}
A.~Roth, C.~Br\"{u}ne, H.~Buhmann, L.~W.~Molenkamp,
J.~Maciejko, X.~L.~Qi, and S.~C.~Zhang, Science \textbf{325},
294 (2009).

\bibitem{wangzhang2008} C.~Liu, T.~L.~Hughes, X.~L.~Qi, K.~Wang, and
  S.~C.~Zhang, Phys. Rev. Lett. \textbf{100}, 236601 (2008).

\bibitem{KaneEtAl2002}
C.~L.~Kane, R.~Mukhopadhyay, and T.~C.~Lubensky,
Phys. Rev. Lett. \tbf{88}, 036401 (2002).

\bibitem{TeoAndKane2014}
J.~C.~Y. Teo and C.~L.~Kane, Phys. Rev. B \tbf{89}, 085101
(2014).

\bibitem{denis} D. Chevallier, P. Szumniak, S.Hoffman, D. Loss, and J. Klinovaja, Phys. Rev. B {\bf 97}, 045404	(2018).

\bibitem{denis2} D. Chevallier and J Klinovaja, Phys. Rev. B {\bf 94}, 035417 (2016).

\bibitem{tip2} M. M. Maska and T. Domanski, Scientific Reports {\bf 7},  16193 (2017).

\bibitem{lin} J. Klinovaja and D. Loss, Phys. Rev. B {\bf 86}, 085408 (2012).

\bibitem{gfactor1} S. B. Zhang, Y. Y. Zhang, and S. Q. Shen, Phys. Rev. B {\bf 90}, 115305 (2014).

\bibitem{gfactor2} R. Skolasinski, D. I. Pikulin, J. Alicea, and M. Wimmer, Phys. Rev. B {\bf 98}, 201404(R) (2018).

\bibitem{gfactor3}  C. Li, S. Zhang, and S. Shen, Phys. Rev. B {\bf 97}, 045420 (2018).

\bibitem{gfactor4} F. Schulz, K. Plekhanov, D. Loss, and J.  Klinovaja, arXiv:2004.10623.

\end{thebibliography}
\end{document}